\documentclass[a4paper,12pt]{article}
\usepackage{amscd,amssymb,amsmath,amsthm}
\usepackage{rotating}
\usepackage{hyperref}
\usepackage{graphicx,caption,dsfont}
\usepackage{mathtools}

\usepackage{subcaption}
\usepackage{amsfonts,enumerate}
\usepackage{fullpage}
\usepackage[numbers,sort & compress]{natbib}
\usepackage{soul}
\usepackage{xfrac}
\usepackage{booktabs}
\numberwithin{equation}{section}
\usepackage{xcolor}

\newtheorem{thm}{Theorem}[section]

\newtheorem{nt}{Note}

\newtheoremstyle{case}{}{}{}{}{}{:}{ }{}
\theoremstyle{case}

\allowdisplaybreaks

\makeatletter
\def\and{%
  \end{tabular}%
  \begin{tabular}[t]{c}}%
\def\@fnsymbol#1{\ensuremath{\ifcase#1\or 1\or 2\or 3\or
   4\or 5\or 6\or 7\or 8\or 9\else\@ctrerr\fi}}
\makeatother
\vspace{-8ex}
\date{}
\begin{document}
\baselineskip=0.70cm
\begin{center}
{\large \textbf{Invariant subspace method for $(m+1)$-dimensional non-linear time-fractional partial differential equations}}
\end{center}
\vspace{-0.5cm}
\begin{center}
P. Prakash$^{1*}$,  K.S. Priyendhu$^{1}$,  M. Lakshmanan$^{2}$\\
$^{1}$Department of Mathematics,\\
Amrita School of Engineering, Coimbatore-641112,\\
  Amrita Vishwa Vidyapeetham, INDIA.\\
$^{2}$Department of Nonlinear Dynamics,\\
 Bharathidasan University,\\
Tiruchirappalli-620 024, INDIA.\\
E-mail ID's: vishnuindia89@gmail.com \&\ p$_{-}$prakash@cb.amrita.edu (P Prakash)\\
\hspace{1.6cm} : priyendhusasikumar@gmail.com (K.S. Priyendhu)\\
\hspace{1.6cm} : lakshman.cnld@gmail.com (M. Lakshmanan)\\
$*$-Corresponding Author
\end{center}
\begin{abstract}
In this paper, we generalize the theory of the invariant subspace method to $(m+1)$-dimensional non-linear time-fractional partial differential equations for the first time. More specifically, the applicability and efficacy of the method have been systematically investigated through the $(3+1)$-dimensional generalized non-linear time-fractional diffusion-convection-wave equation along with appropriate initial conditions. This systematic investigation provides an important technique to find a large class of various types of the invariant subspaces with different dimensions for the above-mentioned equation. Additionally, we have shown that the obtained invariant subspaces help to derive a variety of exact solutions that can be expressed as the combinations of exponential, trigonometric, polynomial and well-known Mittag-Leffler functions.
\end{abstract}
\section{Introduction}
The subject of fractional calculus has drawn much attention and popularity during the recent decades, due to their exact description of various phenomena in a wide variety of fields in science and engineering \cite{pi,kai,kts,vas5,mo11,mn,vas7,vas14,btr,vas2}. In the literature, various types of fractional-order (non-integer order) derivatives and integrals are available that are suggested by Liouville, Riemann, Grunwald, Letnikov, Hadamard, Weyl, Riesz, {Marchaud}, Caputo and other scientists. It is important to observe that the above-mentioned fractional derivatives violate the standard form of Leibnitz rule, chain rule and semigroup properties that are important characteristic properties of fractional-order derivatives. These types of non-standard properties of fractional-order derivatives allow us to describe new and unusual types of complex media and systems \cite{vas2,vas5,vas7}. It is well-known that fractional-order differential operators are non-local and they depend on all the past values of the function and in particular not only on the immediate past. Hence the fractional-order derivatives and integrals are used to add memory to complex systems \cite{vas2}.
Fractional-order derivatives and integrals play a significant role to study the behaviour of various phenomena and systems that are characterized by power-law long term memory, power-law non-locality and fractal properties \cite{vas2,vas5,vas7}. It is well-known that many natural phenomena depend on all the past time history and not only on an instantaneous time. This type of process can be successfully modelled by using the theory of non-integer orders integrals and derivatives \cite{pi,kai,kts,vas5,mo11,mn,vas7,vas14,btr,vas2}.
Also, we would like to point out that the fractional PDEs provide an important way to study the parabolic as well as hyperbolic types of PDEs simultaneously. For instance, we can consider the following time-fractional PDEs (TFPDEs)
\begin{equation}\label{d}
\dfrac{\partial^\alpha u}{\partial t^\alpha}=\gamma\dfrac{\partial^2 u}{\partial x^2},\ t>0,\ \gamma\in\mathbb{R},\ 0<\alpha\leq2,
\end{equation}
which is called a time-fractional diffusion-wave equation \cite{bb1,lk}. Note that equation \eqref{d} describes different phenomena for different values of $\alpha$ that are given below.
\begin{itemize}
\item[$\bullet$] First we note that for $\alpha=1$, equation \eqref{d} is familiarly called the diffusion equation. Also, we observe that equation \eqref{d} becomes the classical wave equation if $\alpha=2$. 
\item[$\bullet$] Also, we note that if $\alpha\in(0,1)$, equation \eqref{d} describes the sub-diffusion phenomenon \cite{bb1,lk} (slow movement of particles) while it represents the super-diffusion phenomenon \cite{bb1,lk} (fast movement of particles) if $\alpha>1$. Additionally, it should be noted that equation \eqref{d} plays a significant role to study the intermediate process between wave and diffusion phenomenon \cite{bb1,lk} if $\alpha\in(1,2)$.
\end{itemize}
 We know that the convection-diffusion equation is governed by the flow of heat or energy in the system when considered under the process of diffusion or convection or both. Models {under} computational simulations giving rise to oceanic waves, models for oil reservoir simulations, groundwater flow, climatic variations and weather predictions, adsorption in a porous medium, viscosity of fluids, population dynamics as well as turbulent heterogeneous physical systems are often represented by fractional PDEs. Due to their rich contributions to applied areas of science and engineering, fractional differential equations (FDEs) have been a focus point of many mathematicians as well as physicists over several years. Some special kinds of equations for the celebrated diffusion-convection equation are the Richards equation \cite{CRD4}, the Burgers equation \cite{CRD4} and the foam drainage equation \cite{CRD4}, to name a few.

The study of FDEs is more complex than their classical (integer-order) differential equations, due to the violation of the Leibnitz rule, semigroup property and chain rule. So the derivation of exact solutions of FDEs is a very tough and challenging task. Due to the above-mentioned reasons, there exist no well-defined methods to analyze the FDEs systematically. We know that finding the exact solutions to FDEs is important because it helps to understand the behaviour and qualitative features of complex phenomena and systems. For this reason, in recent years, many researchers and scientists are engaged in developing certain exact and numerical methods of non-integer order non-linear ODEs and PDEs  such as Lie symmetry method \cite{bb1,lk,praf,kdv,ru4,nas,ppp,CRD2,pra1,pra5}, invariant subspace method \cite{ra,pra4,pra3,ps2,ru2,pa1,s1,pp1,pp2,ru3,pra2,praf}, differential transform method \cite{mo2}, Adomian decomposition method \cite{Ge,mo}, new hybrid method \cite{mma}, variable separation method \cite{rui2022} and so on.

Recent investigations showed that the invariant subspace method is a more efficient and powerful analytical tool to derive exact solutions of integer and non-integer orders scalar and coupled non-linear PDEs. This method is familiarly called the generalized separation of variable method which was initially proposed by Galaktionov and Svirshchevskii \cite{gs} and was further developed by Ma \cite{wx} and many others \cite{wy,wx1,yem,li,zq,ps1} for integer-order non-linear PDEs.In recent years, the invariant subspace method has been extended by Gazizov and Kasatkin \cite{ru2} and others \cite{ra,pra4,pra3,ps2,pa1,s1,pp1,pp2,ru3,pra2,praf,gara1,gara2,IVSM} for scalar and coupled fractional non-linear PDEs. {In 2022, Rui et al \cite{rui2022} have developed the variable separation method for finding the exact solutions of TFPDEs, which is based on the idea of the invariant subspace method.} Very recently, Prakash et al \cite{ppa} and Sangita et al \cite{ps2} have investigated this method for $(2+1)$-dimensional scalar and coupled non-linear TFPDEs.

We wish to point out here that to the best of our knowledge, the above invariant subspace method has not been generalized for $(m+1)$-dimensional ($m>2$) non-linear TFPDEs by anyone. Hence the aim of this work is to show how to generalize the theory of the invariant subspace method for finding the exact solutions of $(m+1)$-dimensional non-linear TFPDEs. In addition, the efficacy and applicability of the method will be investigated through a $(3+1)$-dimensional generalized non-linear time-fractional diffusion-convection-wave (DCW) equation
\begin{equation}\label{cdeqn}
\frac{\partial^{\alpha}u}{\partial t^{\alpha}}=
 \sum\limits^{3}_{r=1}\dfrac{\partial  }{\partial x_r}\left( P_r(u)\frac{\partial u}{\partial x_r}\right)+\sum\limits^{3}_{r=1}Q_r(u)\left(\frac{\partial u}{\partial x_r}\right), \, \alpha\in(0,2]
\end{equation}
along with the appropriate initial conditions,
\begin{itemize}
\item[(i)] $u(x_1,x_2,x_3,0)= \varphi(x_1,x_2,x_3)$  if $0<\alpha\leq1$,
\item[(ii)] $u(x_1,x_2,x_3,0)=\varphi(x_1,x_2,x_3)$ \& ${\dfrac{\partial u}{\partial t}}\big{|}_{t=0}=\phi(x_1,x_2,x_3)$ if $1<\alpha\leq2$,
\end{itemize}
where $t>0,x_r\in \mathbb{R},r=1,2,3, u=u(x_1,x_2,x_3,t)$ and $ P_r(u) ,Q_r(u) ,r=1,2,3,$ denote the diffusion process and convection process respectively and $\dfrac{\partial^{\alpha}u}{\partial t^{\alpha}} $ denotes the Caputo fractional derivative of $u$ with respect to $t$ of order $\alpha$, which is defined \cite{pi,kai} as
\begin{eqnarray}
\begin{aligned}\label{c}
\dfrac{\partial^\alpha u}{\partial t^\alpha}=
\left\{ \begin{array}{ll}
\dfrac{1}{\Gamma(n-\alpha)}\int\limits_{0}^{t}\dfrac{\partial ^{n} u(x_1,x_2,\tau)}{\partial \tau^{n} }(t-\tau)^{n-\alpha-1}d\tau, & n-1<\alpha< n ,\ n\in\mathbb{N}; \\
\dfrac{\partial^ n u}{\partial t^n}, & \alpha=n \in\mathbb{N}.
 \end{array}
 \right.
\end{aligned}
\end{eqnarray}
In this work, we are considering essentially the Caputo fractional derivative.
This article gives a systematic analysis to find the invariant subspace for the $(3+1)$-dimensional (higher dimensional) non-linear time-fractional DCW equation. In addition, we show that using the obtained invariant subspaces, how to find the exact solutions of \eqref{cdeqn} along with the appropriate initial conditions.

The article is organised as follows: Section 2 explains how to generalize the invariant subspace method to find various types of invariant subspaces admitted by $(3+1)$-dimensional non-linear TFPDEs. Also, we extend the invariant spaces to $(4+1)$-dimensional non-linear TFPDEs in section 3.
 Section 4 provides a detailed and systematic analysis to generalize the theory of the invariant subspace method for the $(m+1)$-dimensional non-linear TFPDEs. The efficacy and applicability of the method have been investigated to find various types of invariant subspaces through the $(3+1)$-dimensional DCW equation in section 5. Using the obtained invariant subspaces, we present how to construct various types of exact solutions for the $(3+1)$-dimensional non-linear time-fractional DCW equation \eqref{cdeqn} along with appropriate initial conditions in section 6. Finally, section 7 provides a brief discussion on the applicability of the invariant subspace method to several classical diffusion type equations and their modifications. Also, the concluding remarks of the present work are given in brevity.
\section{Generalized $(3+1)$-dimensional non-linear TFPDEs: Invariant subspace method}
In this section, we explain how to find systematically the various types of invariant subspaces for the following generalized non-linear partial differential operator involving three independent variables as follows
\begin{eqnarray}\begin{aligned}\label{3dop}
\mathcal{\hat{N}}_1[u]=\mathcal{\hat{N}}_1\left( x_{1},x_{2},x_{3},u,
\frac{\partial u}{\partial x_1},\frac{\partial u}{\partial x_2},\frac{\partial u}{\partial x_3},\ldots,\frac{\partial^k u}{\partial x_1^k},\frac{\partial^k u}{\partial x_2^k},\frac{\partial^k u}{\partial x_3^k},
\frac{\partial^{2}u}{\partial x_1\partial x_2},\ldots,\frac{\partial^{k}u}{\partial x_1^{k_1}\partial x_2^{k_2}{\partial x_3^{k_3}}}\right),
\end{aligned}
\end{eqnarray}
where $t>0$, $x_i\in\mathbb{R},i=1,2,3,$ $u=u(x_1,x_2,x_3,t)$ and the given non-linear partial differential operator $\mathcal{\hat{N}}_1[u]$ of order $k$ is the sufficiently smooth and $k=\sum\limits_{i=1}^{3}k_i$, $k_i \in \mathbb{N},i=1,2,3$. Additionally, using the obtained invariant subspaces, we present how to find the exact solutions for the $(3+1)$-dimensional non-linear TFPDEs.
First, we define the finite dimensional linear space ${_{3}\mathcal{\hat{W}}_n}$ as
\begin{equation}
\label{3dls}
{_3\mathcal{\hat{W}}_n}=\left\{\sum\limits_{q=1}^{n}\lambda_q\xi_q(x_1,x_2,x_3)\big{|} \xi_q(x_1,x_2,x_3)\in \mathcal{B}, \lambda_q \in \mathbb{R},q=1,2,\dots,n\right\},
\end{equation}
where the basis set $\mathcal{B}$ is given by
$\mathcal{B}=\left\{ \xi_1(x_1,x_2,x_3),\xi_2(x_1,x_2,x_3),\dots,\xi_n(x_1,x_2,x_3)\right\}$
for some linearly independent functions $\xi_q(x_1,x_2,x_3)\; (q=1,2,\dots,n)$ and $n$ is the dimension of the linear space ${_{3}\mathcal{\hat{W}}_n}$, that is $\text{dim}({_{3}\mathcal{\hat{W}}_n})=n$ $(<\infty)$.

 A given finite dimensional linear space ${_3\mathcal{\hat{W}}_n}$ is invariant under the given partial differential operator $\mathcal{\hat{N}}_1[u]$ if $ \mathcal{\hat{N}}_1[u] \in {_3\mathcal{\hat{W}}_n}$ whenever $ u\in{_3\mathcal{\hat{W}}_n}.$
If ${_3\mathcal{\hat{W}}_n}$ is invariant under $\mathcal{\hat{N}}_1[u]$, then we obtain
\begin{equation}\label{3dinv}
\mathcal{\hat{N}}_1[\sum\limits_{q=1}^{n}\lambda_q\xi_q(x_1,x_2,x_3)]=\sum\limits_{q=1}^{n}\Lambda_q(\lambda_1,\lambda_2,\dots,\lambda_n)\xi_q(x_1,x_2,x_3),
\end{equation}
where the functions $\Lambda_q(\lambda_1,\lambda_2,\dots,\lambda_n) ,q=1,2,\dots,n,$ are coefficients of expansion with respect to $\mathcal{B}.$
\begin{thm}\label{thm1}
Consider the generalized $(3+1)$-dimensional non-linear TFPDE
\begin{equation}\label{3dpde}
\frac{\partial^{\alpha}u}{\partial t^{\alpha}}=\mathcal{\hat{N}}_1[u],\ t>0,\ \alpha>0,
\end{equation}
where $u=u(x_1,x_2,x_3,t)$. 
 If the above non-linear partial differential operator $\mathcal{\hat{N}}_1[u]$ preserves the invariant subspace  ${_3\mathcal{\hat{W}}_n}$,  then there exists an exact solution to the equation \eqref{3dpde} of the form 
\begin{equation} \label{3dsol}
{u}(x_1,x_2,x_3,t)=\Psi_1(t)\xi_1(x_1,x_2,x_3)+\Psi_2(t)\xi_2(x_1,x_2,x_3)+\dots+\Psi_n(t)\xi_n(x_1,x_2,x_3),
\end{equation}
where the functions $\Psi_q(t)\ (q=1,2,\dots,n)$ satisfy the system of fractional-order ODEs
\begin{equation}\label{3dsys}
\dfrac{d^{\alpha}\Psi_q(t)}{dt^\alpha}=\Lambda_q(\Psi_1(t),\Psi_2(t),\dots,\Psi_n(t)),\; q=1,2,\dots,n.
\end{equation}
\end{thm}
\textbf{Proof:}
Let us assume that the non-linear operator $\mathcal{\hat{N}}_1[u]$ preserves the invariant subspace  ${_3\mathcal{\hat{W}}_n}$ given in \eqref{3dls}.
Thus, by using equation \eqref{3dinv}, we obtain
\begin{equation}\label{3d1}
\mathcal{\hat{N}}_1[u(x_1,x_2,x_3,t)] =\sum \limits^{n} _{q=1}\Lambda_q(\Psi_1(t),\Psi_2(t),...,\Psi_n(t))\xi_q(x_1,x_2,x_3).
\end{equation}
Computing the Caputo fractional derivative with respect to $t$ of order $\alpha>0$ of the function
${u}(x_1,x_2,x_3,t)=\sum \limits_{q=1}^n \Psi_q(t)\xi_q(x_1,x_2,x_3)$, we have
\begin{equation}\label{3d2}
 \frac{\partial^{\alpha }{u}}{\partial t^{\alpha}}=\sum^{n} _{q=1}  \dfrac{d^{\alpha}\Psi_q(t)}{dt^\alpha}\xi_q(x_1,x_2,x_3).
 \end{equation}
Using the equations \eqref{3d1} and \eqref{3d2} in \eqref{3dpde}, we obtain the following
\begin{eqnarray}\begin{aligned}\label{3d3}
&\sum\limits_{q=1}^{n}\left[\dfrac{d^{\alpha}\Psi_q(t)}{dt^\alpha}-\Lambda _q(\Psi_1(t),\Psi_2(t),\dots,\Psi_n(t))\right]\xi_q(x_1,x_2,x_3)=0,
\end{aligned}
\end{eqnarray}
where $\xi_q(x_1,x_2,x_3)$ $(q=1,2,\dots,n)$ are linearly independent functions which form a basis for the linear space ${_3\mathcal{\hat{W}}_n}$ given in \eqref{3dls}. Thus, we have
\begin{eqnarray}\begin{aligned}
\dfrac{d^{\alpha}\Psi_q(t)}{dt^\alpha}=\Lambda_q(\Psi_1(t),\Psi_2(t),\dots,\Psi_n(t)),\; q=1,2,\dots,n.
\end{aligned}
\end{eqnarray}
\begin{nt}
It should be noted that the invariant subspace method allows us to reduce a large class of non-linear TFPDEs into systems of arbitrary-order ODEs.
If the obtained system of non-integer order ODEs may be solved by well-known analytical methods, then the obtained exact solutions can be viewed in the framework of finite variable separable structure.
\end{nt}
Now, we explain how to define various types of invariant subspaces for the given non-linear differential operator $\mathcal{\hat{N}}_1[\hat{u}]$.
\subsection{Invariant subspaces preserved by the given differential operator \eqref{3dop}}
Now, let us consider the linear homogeneous ODEs of order $n_i$ in the form
\begin{equation}\label{3d01}
\mathcal{D}_{x_i}[\omega]\equiv \dfrac{d^{n_i}\omega}{d x_{i}^{n_i}}+ a_{in_i-1}\dfrac{d^{n_i-1}\omega}{d x_{i}^{n_i-1}}+\cdots+a_{i0}\omega=0,\ i=1,2,3
\end{equation}
with the set of $n_i$ linearly independent solutions as ${_3 \mathcal{S}_{x_i}}=\{\omega_{q_{i}}^{i}(x_i)\big{|}q_i=1,2,\dots,n_i\} $, $ a_{ij}\in \mathbb{R},$ $j=0,1,\dots,n_i-1,\, i=1,2,3.$
Thus, we can write the solution spaces of the given linear homogeneous ODEs \eqref{3d01} as follows,
\begin{equation}
\label{3dsspace}
\mathcal{\hat{W}}_{n_i}^{x_i}=\text{Span}({_3\mathcal{S}_{x_i}})=\left\{\sum_{q=1}^{n_i}\lambda_{q_{i}}\omega_{q_{i}}^{i}(x_i)\big{|}\mathcal{D}_{x_i}[\omega_{q_{i}}^{i}(x_i)]=0,\lambda_{q_{i}}\in \mathbb{R},q_i=1,2,\dots,n_{i}\right\}.
\end{equation}
Now, we can define three types of invariant subspaces for the non-linear partial differential operator $\mathcal{\hat{N}}_1[u]$ given in \eqref{3dop} that are discussed below in detail.\\
\textbf{Type-1 invariant subspace} $ {_3^{1}\mathcal{\hat{W}}}_{l} $ : Let us first construct the Type-1 linear space for the operator $\mathcal{\hat{N}}_1[u]$ as
\begin{eqnarray}\begin{aligned}\label{3dt1}
{_3^{1}\mathcal{\hat{W}}}_{l}=&\text{Span}({_3\mathcal{\hat{S}}}^{1})
=\left\{\xi(x_1,x_2,x_3)=\sum_{q_{1},q_{2},q_3}\lambda_{q_{1},q_{2}, q_{3}}\prod_{i=1}^{3}\omega_{q_{i}}^{i}(x_{i})  \big{|}\mathcal{D}_{x_i}[\xi(x_1,x_2,x_3)]=0,\right.\\
&\left.
\lambda_{q_{1},q_{2}, q_{3}}\in \mathbb{R},q_{i}=1,2,\dots,n_{i},i=1,2,3\right. \Biggr\},
\end{aligned}
\end{eqnarray}
where the basis set 
$ {_3\mathcal{\hat{S}}^1}=\left\{\prod\limits_{i=1}^{3}\omega_{q_i}^i(x_i)\big{|}\omega_{q_i}^i(x_i)\in {_1\mathcal{\hat{S}}}_{x_i},\,q_i=1,2,\ldots,n_i,i=1,2,3\right\}$
and $l$ denotes the dimension of Type-1 linear space ${_3^{1}\mathcal{\hat{W}}}_{l}$, that is $\text{dim}({_3^{1}\mathcal{\hat{W}}}_{l})=l= n_1n_2n_3$.\\
The Type-1 linear space \eqref{3dt1} is invariant under the given differential operator $\mathcal{\hat{N}}_1[u]$ if $\mathcal{\hat{N}}_1[u]\in {_3^{1}\mathcal{\hat{W}}}_{l}$ whenever $ u\in {_3^{1}\mathcal{\hat{W}}}_{l}$. Thus, the obtained invariant criteria for the Type-1 linear space ${_3^{1}\mathcal{\hat{W}}}_{l}$ are in the following form 
 \begin{eqnarray}\begin{aligned}\label{invc3dt1}
 &\mathcal{D}_{x_i}\left[\mathcal{\hat{N}}_1[u]\right]=0,\, \text{whenever} \,
 &\mathcal{D}_{x_i}[u]=0,\; i=1,2,3
 \end{aligned}\end{eqnarray}
 along with the differential consequences with respect to $x_i,i=1,2,3.$
\\
\textbf{Type-2 invariant subspace} $ {_3^{2}\mathcal{\hat{W}}}_{l} $ :  Now, let $\omega_{n_{r}}^r(x_{r})=1\in {_3\mathcal{{S}}_{x_{r}}},r=1,2,3$. For this case, we can construct the Type-2 invariant subspace $ {_3^{2}\mathcal{\hat{W}}}_{l} $ for the given operator $\mathcal{\hat{N}}_1[u]$ as
\begin{eqnarray}\begin{aligned}\label{3dt2}
{_3^{2}\mathcal{\hat{W}}}_{l}=&\text{Span}({_3\mathcal{\hat{S}}^{2}})
=\left\{\xi(x_1,x_2,x_3)=\lambda+\sum_{i=1}^3\sum_{q_{i}=1}^{n_i-1}\lambda_{q_{i}}^{i}\omega_{q_{i}}^{i}(x_{i})+\sum_{\substack{i<j\\r =1\\r=i,j }}^{3}\sum_{\substack{q_{r}=1,\\r=i,j}}^{{n_r-1}}\lambda_{q_{i},q_{j}}^{i,j}\omega_{q_{i}}^{i}(x_{i}) \omega_{q_{j}}^{j}(x_{j})\big{|}\right.\\
&\left. \mathcal{D}_{x_{i}}[\xi(x_1,x_2,x_3)]=0,
a_{i0}=0,
\lambda_{q_{i},q_{j} }^{i,j},\lambda_{q_{i}}^{i},\lambda\in \mathbb{R},q_{r}=1,2,\dots,n_{r}-1,\right.\\
&\left.
r=i,j=1,2,3,\dfrac{\partial^3 \xi(x_1,x_2,x_3) }{\partial x_1\partial x_2\partial x_3}=0\right.\Biggr\},
\end{aligned}
\end{eqnarray}
where the basis set
${_3\mathcal{\hat{S}}^{2}}=\bigcup\limits_{i<j}\left\{\omega_{q_{i}}^{i}(x_{i})\omega_{q_{j}}^{j}(x_{j})\big{|} \omega_{q_{r}}^r(x_{r})\in \mathcal{{S}}_{x_{r}},
q_{r}=1,\ldots,n_{r},r=i,j=1,2,3\right\}
$ and the
dimension of the Type-2 invariant subspace
\[ \text{dim}({_3^{2}\mathcal{\hat{W}}}_{l})=l=1+\sum\limits _{i =1}^{3}(n_i-1)\ +\sum\limits _{\substack{i<j\\i,j =1,2,3 }}(n_i-1)(n_j-1).\]
Thus, the invariant conditions for Type-2 invariant subspace ${_3^{2}\mathcal{\hat{W}}}_{l}$ read in the form
\begin{eqnarray}\begin{aligned}\label{invc3dt2}
&\mathcal{D}_{x_{i}}\left[\mathcal{\hat{N}}_1[u]\right]=0,a_{i0}=0,
\dfrac{\partial^3 \mathcal{\hat{N}}[u] }{\partial x_1\partial x_2\partial x_3}=0,\text{if}\,
\mathcal{D}_{x_{i}}[u]=0,a_{i0}=0,
\dfrac{\partial^3 u }{\partial x_1\partial x_2\partial x_3}=0
\end{aligned}
\end{eqnarray}
along with the differential consequences with respect to $x_i,i=1,2,3.$
\\
\textbf{Type-3 invariant subspace} $ {_3^{3}\mathcal{\hat{W}}}_{l} $ : Finally, let us assume that $\omega_{q_{r}}^r(x_{r})\in{_3\mathcal{{S}}_{x_{r}}}, q_r=1,2,\dots,n_r-1$ and $\omega_{n_{r}}^r(x_{r})=1\in {_3\mathcal{{S}}_{x_{r}}},r=1,2,3.$
Thus, we define the Type-3 invariant subspace for the given differential operator $\mathcal{\hat{N}}_1[u]$ as
 \begin{eqnarray}\begin{aligned}\label{3dt3}
{_3^{3}\mathcal{\hat{W}}}_{l}=&\text{Span}({_3\mathcal{\hat{S}}}^{3})
= \left\{\xi(x_1,x_2,x_3)= \lambda+\sum_{i=1}^{3}\lambda_{q_i}\omega_{q_i}^{i}(x_i)\big{|}\mathcal{D}_{x_i}[\xi(x_1,x_2,x_3)]=0,a_{i0}=0,
\right.\\&\left.\dfrac{\partial^2\xi(x_1,x_2,x_3)}{\partial x_{i}\partial x_{j}}=0,i<j,\lambda,\lambda_{q_{i}}\in \mathbb{R},
q_{i}=1,\dots,n_{i}-1,i,j=1,2,3\right.\Biggr\},
\end{aligned}
\end{eqnarray}
where the basis set
\[ {_3\mathcal{\hat{S}}^{3}}=\left\{1,\omega_{1}^{1}(x_1),\dots,\omega_{n_{1}-1}^{1}(x_1),\omega_{1}^{2}(x_2),\dots,\omega_{n_{2}-1}^{2}(x_2),\omega_{1}^{3}(x_3),\dots,\omega_{n_{3}-1}^{3}(x_3)\right\}
\]
and the dimension of the linear space
$\text{dim}({_3^{3}\mathcal{\hat{W}}}_{l})=l=n_1+n_{2}+n_3-2.$\\
In this case, the invariance conditions read in the form
\begin{eqnarray}\begin{aligned}\label{invc3dt3}
&\mathcal{D}_{x_i}\left[\mathcal{\hat{N}}_1[u]\right]=0,a_{i0}=0,
\dfrac{\partial^2\mathcal{\hat{N}}_1[u]}{\partial x_{i}\partial x_{j}}=0,
\text{if}\,
\mathcal{D}_{x_i}[u]=0,a_{i0}=0,
\dfrac{\partial^2u}{\partial x_{i}\partial x_{j}}=0
\end{aligned}
\end{eqnarray}
along with the differential consequences with respect to $x_i,$ where $i,j=1,2,3.$
\begin{nt}
It is interesting to note that
if $1 \in {_3^{1}\mathcal{\hat{W}}}_{l}$, then we obtain the following inclusion relation
 \[{_3^{3}\mathcal{\hat{W}}}_{l}\subseteq{_3^{2}\mathcal{\hat{W}}}_{l}\subseteq{_3^{1}\mathcal{\hat{W}}}_{l}. \]
 From this, we observe that ${_3^{j} \mathcal{\hat{W}}}_l $ is invariant with respect to the given differential operator $\mathcal{\hat{N}}_1[u]$, but the linear space $ {_3^{i} \mathcal{\hat{W}}}_l $ need not be an invariant under $\mathcal{\hat{N}}_1[u]$, where  $i<j,\ (i,j=1,2,3)$.\end{nt}
Next, we give the detailed systematic analysis for constructing various types of linear spaces for the generalized $(4+1)$-dimensional non-linear TFPDEs.
\section{Generalized $(4+1)$-dimensional non-linear TFPDEs : Invariant subspace method}
In this section, we consider the generalized non-linear partial differential operator involving four independent variables in the form
\begin{eqnarray}\begin{aligned}\label{4dop}
\mathcal{\hat{N}}_2[u]\equiv&\mathcal{\hat{N}}_2\left( x_{1},\dots,x_4,u,
\frac{\partial u}{\partial x_1},\dots,\frac{\partial u}{\partial x_4},\dots,\frac{\partial^k u}{\partial x_1^k},\dots,
\frac{\partial^{2}u}{\partial x_1\partial x_2},\dots,\frac{\partial^{k}u}{\partial x_1^{k_1}\partial x_2^{k_2}\partial x_3^{k_3}\partial x_4^{k_4}}\right),\end{aligned}
\end{eqnarray}
where $u=u(x_1,x_2,x_3,x_4,t)$, $x_i\in\mathbb{R}\ , i=1,2,3,4$, $t>0$ and the non-linear partial differential operator $\mathcal{\hat{N}}_2[u]$ of order $k$ is the given sufficiently smooth operator and $\sum\limits_{i=1}^{4}k_i=k$, $k_i \in \mathbb{N},i=1,2,3,4$.\\
Now, we define the finite dimensional linear space
 \begin{equation}
\label{4dls}
{_4\mathcal{\hat{W}}_n}=\left\{\sum\limits_{q=1}^{n}\lambda_q\xi_q(x_1,x_2,x_3,x_4)\big{|} \xi_q(x_1,x_2,x_3,x_4)\in \mathcal{B}, \lambda_q \in \mathbb{R},q=1,2,\dots,n\right\},
\end{equation}
where $n$ ($<\infty$) is the dimension of the linear space ${_4\mathcal{\hat{W}}_n}$ and the basis set $\mathcal{B}$ for the given linear space \eqref{4dls} is
$
 \mathcal{B}=\left\{ \xi_q(x_1,x_2,x_3,x_4) | q=1,2,\dots,n\right\}
$
for some linearly independent functions  $ \xi_q(x_1,x_2,x_3,x_4) , q=1,2,\dots,n $.

A given finite dimensional linear space ${_4\mathcal{\hat{W}}_n}$ is invariant under $\mathcal{\hat{N}}_2[u] $ if $\mathcal{\hat{N}}_2[u] \in{_4\mathcal{\hat{W}}_n}$ whenever $ u\in{_4\mathcal{\hat{W}}_n}. $
If the operator $\mathcal{\hat{N}}_2[u] $ preserves the linear space ${_4\mathcal{\hat{W}}_n}$, then we have
\begin{equation}\label{4dinv}
\mathcal{\hat{N}}_2[\sum\limits_{q=1}^{n}\lambda_q\xi_q(x_1,x_2,x_3,x_4)]=\sum\limits_{q=1}^{n}\Lambda_q(\lambda_1,\lambda_2,\dots,\lambda_n)\xi_q(x_1,x_2,x_3,x_4),
\end{equation}
where the functions $\Lambda_q(\lambda_1,\lambda_2,\dots,\lambda_n),q =1,2,\dots,n,$ are the coefficients of expansion with respect to $\mathcal{B}.$
\begin{thm}
Consider the generalized $(4+1)$-dimensional non-linear TFPDE
\begin{equation}\label{4dpde}
\frac{\partial^{\alpha}u}{\partial t^{\alpha}}=\mathcal{\hat{N}}_2[u],\ t>0,\ \alpha>0,
\end{equation}
where $u=u(x_1,x_2,x_3,x_4,t)$. 
If the linear space ${_4\mathcal{\hat{W}}_n}$ given in \eqref{4dls} is invariant under the given operator $\mathcal{\hat{N}}_2[u]$, then there exists an exact solution to the equation \eqref{4dpde} of the form
\begin{equation} \label{4dsol}
{u}(x_1,x_2,x_3,x_4,t)=\sum _{q=1}^n \Psi_q(t)\xi_q(x_1,x_2,x_3,x_4),
\end{equation}
where the functions $\Psi_q(t),q=1,2,\dots,n $ satisfy the system of arbitrary-order ODEs
\begin{equation}\label{4dsys}
\dfrac{d^{\alpha}\Psi_q(t)}{dt^\alpha}=\Lambda_q(\Psi_1(t),\Psi_2(t),\dots,\Psi_n(t)),\; q=1,2,\dots,n.
\end{equation}
\end{thm}
\textbf{Proof:} Similar to the proof of theorem \ref{thm1}.

Next, we give a systematic analysis for defining various types of invariant subspaces determined by the given operator $ \mathcal{\hat{N}}_2[\hat{u}] $ as an extension of types of invariant subspaces for $(3+1)$-dimensional non-linear TFPDEs \eqref{3dpde}.
\subsection{Invariant subspaces admitted by the given differential operator \eqref{4dop}}
Consider the linear homogeneous ODEs of order $n_i$ of the form
\begin{equation}\label{4d01}
\mathcal{D}_{x_i}[\omega]\equiv \dfrac{d^{n_i}\omega}{d x_{i}^{n_i}}+ a_{in_i-1}\dfrac{d^{n_i-1}\omega}{d x_{i}^{n_i-1}}+\cdots+a_{i0}\omega=0,i=1,2,3,4
\end{equation}
with the set of $n_i$ linearly independent solutions as $ {_{4}\mathcal{S}_{x_i}}=\{\omega_{q_{i}}^{i}(x_i)\big{|}q_i=1,2,\dots,n_i\} $ and $ a_{ij} \in \mathbb{R},$ $j=0,1,\dots,n_i-1,\, i=1,2,3,4.$
Solution spaces of the given linear homogeneous ODEs \eqref{4d01} are defined as
\begin{equation}
\label{4dsspace}
\mathcal{\hat{W}}^{x_i}_{n_i}=\text{Span}({_{4}\mathcal{S}_{x_i}})=\left\{\sum_{q=1}^{n_i}\lambda_{q_{i}}\omega_{q_{i}}^{i}(x_i)\big{|}\mathcal{D}_{x_i}[\omega_{q_{i}}^{i}(x_i)]=0,q_i=1,2,\dots,n_{i}\right\}.
\end{equation}
Now, we explain how to define four types of linear spaces for the given non-linear partial differential operator $\mathcal{\hat{N}}_2[u]$.\\
\textbf{Type-1 invariant subspace} $ {^{1}_{4} \mathcal{\hat{W}}}_l $: Let us first define the Type-1 linear subspace $ {^{1}_{4} \mathcal{\hat{W}}}_l $ for the given operator $\mathcal{\hat{N}}_2[u]$ as
\begin{eqnarray}\begin{aligned}\label{4dt1}
{^{1}_{4} \mathcal{\hat{W}}}_l =&\text{Span}({_{4}{\mathcal{\hat{S}}}^{1}} )
=\left\{\xi(x_1,x_2,x_3,x_4)=\sum_{q_{1},\dots,q_4}\lambda_{q_{1},q_{2}, q_{3},q_4}\prod_{i=1}^{4}\omega_{q_{i}}^{i}(x_{i})  \big{|}
\right.\\
&\left.\mathcal{D}_{x_i}[\xi(x_1,x_2,x_3,x_4)]=0,
\lambda_{q_{1},q_{2}, q_{3},q_4}\in \mathbb{R},q_{i}=1,2,\dots,n_{i},i=1,2,3,4\right. \Biggr\},
\end{aligned}
\end{eqnarray}
where
the set $ {_{4}\mathcal{\hat{S}}^{1}}=\left\{\prod\limits_{i=1}^{4}\omega_{q_i}^{i}(x_i)\big{|} \; \omega_{q_i}^{i}(x_i)\in {_4\mathcal{\hat{S}}_{x_i}},\,q_i=1,2,\ldots,n_i,i=1,2,3,4\right\} $
and $l$ denotes the dimension of the space ${^{1}_{4} \mathcal{\hat{W}}}_{l}$, that is
 $ \text{dim}({^{1}_{4} \mathcal{\hat{W}}} _l)=l= n_1n_2n_3n_4. $\\
The Type-1 linear space ${^{1}_{4} \mathcal{\hat{W}}}_l$ given in \eqref{4dt1} is invariant under the given non-linear partial differential operator $ \mathcal{\hat{N}}_2[u]$ if $\mathcal{\hat{N}}_2[u]\in {^{1}_4 \mathcal{\hat{W}}}_l $ whenever $ u\in {^{1}_4 \mathcal{\hat{W}}} _l$. Thus, the invariance conditions for Type-1 invariant subspace ${^{1}_4 \mathcal{\hat{W}}}_l $ can be read as
 \begin{eqnarray}\begin{aligned}\label{invc4dt1}
 &\mathcal{D}_{x_i}\left[\mathcal{\hat{N}}_2[u]\right]=0
 \,\, \text{whenever}\,\,\mathcal{D}_{x_i}[u]=0,\; i=1,2,3,4 ,
 \end{aligned}\end{eqnarray}
 along with the differential consequences with respect to $x_i,i=1,2,3,4.$
 \\
\textbf{Type-2 invariant subspace} ${^{2}_{4}\mathcal{\hat{W}}}_{l}  $ :
 The Type-2 linear space for the given non-linear partial differential operator $\mathcal{\hat{N}}_2[u]$ is defined in the following form
 \begin{eqnarray}\begin{aligned}\label{4dt2}
{^{2}_{4}\mathcal{\hat{W}}}_{l}=&\text{Span}({_{4}\mathcal{{\hat{S}}}^{2}})
=\left\{\xi(x_1,x_2,x_3,x_4)=\lambda+\sum_{i=1}^4\sum_{q_{i}=1}^{n_i-1}\lambda_{q_{i}}^{i}\omega_{q_{i}}^{i}(x_{i})
+\sum_{\substack{i<j}}\sum_{\substack{q_{i},q_{j}}}\lambda_{q_{i},q_{j}}^{i,j}\omega_{q_{i}}^{i}(x_{i}) \omega_{q_{j}}^{j}(x_{j})
\right.\\
&\left.
+\sum_{\substack{i<j<s}}\sum_{\substack{q_{i},q_{j},q_{s}}}^{{n_r-1}}\lambda_{q_{i},q_{j},q_{s}}^{i,j,s}\omega_{q_{i}}^{i}(x_{i}) \omega_{q_{j}}^{j}(x_{j}) \omega_{q_s}^{s}(x_{s})\,\,\big{|}\,\,
\lambda_{q_{i},q_{j}}^{i,j},\lambda_{q_{i},q_{j},q_{s} }^{i,j,s},\lambda_{q_{i}}^{i},\lambda\in \mathbb{R},
\right.\\
&\left.
\mathcal{D}_{x_{i}}[\xi]=0,a_{i0}=0,
\dfrac{\partial^4 \xi }{\partial x_1\partial x_2\partial x_3\partial x_4}=0,
q_{r}=1,2,\dots,n_{r}-1,
r=i,j,s=1,2,3,4\right.\Biggr\},
\end{aligned}
\end{eqnarray}
where the basis set
\[{_{4}\mathcal{\hat{S}}^{2}}=\bigcup_{i<j}\left\{\omega_{q_{i}}^{i}(x_{i})\omega_{q_{j}}^{j}(x_{j}) \omega_{q_s}^{s}(x_{s})\big{|} \omega_{q_{r}}^r(x_{r})\in {_4\mathcal{{S}}_{x_{r}}},
q_{r}=1,2,\dots,n_{r},r=i,j,s=1,2,3,4\right\}
\] such that $ \omega_{n_{r}}^r(x_{r})=1\in {_{4}\mathcal{{S}}_{x_{r}}},\ r=1,2,3,4,$
and the dimension of the linear space ${^{2}_{4}\mathcal{\hat{W}}}_{l}$ obtained as
\[ \text{dim}({^{2}_{4}\mathcal{\hat{W}}}_{l})=1+\sum_{i=1}^{4}(n_i-1)+\sum\limits _{\mathclap{\substack{i<j\\i,j =1,2,3,4 }}}(n_i-1)(n_j-1)+
\sum\limits _{\mathclap{\substack{i<j<s\\i,j,s =1,2,3,4 }}}(n_i-1)(n_j-1)(n_s-1).\]
The Type-2 linear space ${^{2}_{4}\mathcal{\hat{W}}}_{l}$ is invariant under the given operator $\mathcal{\hat{N}}_2[u]$ if 
$ \mathcal{\hat{N}}_2[u]\in {^{2}_{4}\mathcal{\hat{W}}}_{l}$
whenever $ u\in {^{2}_{4}\mathcal{\hat{W}}}_{l}. $
 Thus, the invariance conditions for the Type-2 linear space ${^{2}_{4}\mathcal{\hat{W}}}_{l}$ are obtained as follows
\begin{eqnarray*}\begin{aligned}\label{invc4dt2}
&\mathcal{D}_{x_{i}}\left[\mathcal{\hat{N}}_2[u]\right]=0,
\dfrac{\partial^4 \mathcal{\hat{N}}_2[u] }{\partial x_1\partial x_2\partial x_3\partial x_4}=0,a_{i0}=0
\,\text{if}\,
\mathcal{D}_{x_{i}}[u]=0,
\dfrac{\partial^4 u}{\partial x_1\partial x_2\partial x_3\partial x_4}=0,a_{i0}=0,\\
\end{aligned}
\end{eqnarray*}
along with the differential consequences with respect to $x_i,i=1,2,3,4.$
\\
\textbf{Type-3 invariant subspace} $ {^{3}_{4}\mathcal{\hat{W}}}_{l} $:
Now, let $\omega_{n_{r}}^r(x_{r})=1\in {_4\mathcal{{S}}_{x_{r}}}$ and we define the Type-3 linear space for the given operator $\mathcal{\hat{N}}_2[u]$ as 
\begin{eqnarray}\begin{aligned}\label{4dt3}
{^{3}_{4}\mathcal{\hat{W}}}_{l}&=\text{Span}({_{4}\mathcal{\hat{S}}}^{3})
=\left\{\xi(x_1,x_2,x_3,x_4)=\lambda+\sum_{i=1}^4\sum_{q_{i}=1}^{n_i-1}\lambda_{q_{i}}^{i}\omega_{q_{i}}^{i}(x_{i})
\right.\\
&\left.
+\sum_{\substack{i<j}}\sum_{\substack{q_{i},q_{j}}}\lambda_{q_{i},q_{j}}^{i,j}\omega_{q_{i}}^{i}(x_{i}) \omega_{q_{j}}^{j}(x_{j})
\,\big{|}
\lambda_{q_{i},q_{j}}^{i,j},\lambda_{q_{i}}^{i},\lambda\in \mathbb{R},
\mathcal{D}_{x_{i}}[\xi]=0,a_{i0}=0,\right.\\
&\left.
\dfrac{\partial^4 \xi }{\partial x_i\partial x_j\partial x_s}=0,
q_{r}=1,2,\dots,n_{r}-1,i<j<s,
i,j,s=1,2,3,4\right.\Biggr\},
\end{aligned}
\end{eqnarray}
where the dimension of ${^{3}_{4}\mathcal{\hat{W}}}_{l}$ is
$\text{dim}({^{3}_{4}\mathcal{\hat{W}}}_{l})=1+\sum\limits_{i=1}^{4}(n_i-1)+\sum\limits _{\mathclap{\substack{i<j}}}(n_i-1)(n_j-1),i,j =1,2,3,4,$
and the basis set ${_{4}\mathcal{\hat{S}}}^{3}$ for the linear space \eqref{4dt3} is
\[{_{4}\mathcal{\hat{S}}}^{3}=\bigcup\limits_{i<j}\left\{\omega_{q_{i}}^{i}(x_{i})\omega_{q_{j}}^{j}(x_{j})\big{|}
  \omega_{q_{r}}^r(x_{r})\in {_4\mathcal{{S}}_{x_{r}}},
q_{r}=1,\dots,n_{r},
r=i,j=1,\dots,4\right\}.\]
The Type-3 linear space ${^{3}_{4}\mathcal{\hat{W}}}_{l}$ is invariant under $\mathcal{\hat{N}}_2[u]$ if 
$ \mathcal{\hat{N}}_2[u]\in {^{3}_4\mathcal{\hat{W}}}_{l}$
whenever $ u\in {^{3}_4\mathcal{\hat{W}}}_{l}.$
Hence the invariance conditions for the Type-3 invariant subspace ${^{3}_{4}\mathcal{\hat{W}}}_{l}$ read as
\begin{eqnarray*}\begin{aligned}\label{invc4dt3}
&\mathcal{D}_{x_{i}}\left[\mathcal{\hat{N}}_2[u]\right]=0,a_{i0}=0,
\dfrac{\partial^3 \mathcal{\hat{N}}_2[u] }{\partial x_i\partial x_j\partial x_s}=0,i<j<s, i,j,s=1,2,3,4,
\\ \text{if} \, \,&
\mathcal{D}_{x_{i}}[u]=0,a_{i0}=0,
\dfrac{\partial^3 u}{\partial x_i\partial x_j\partial x_s}=0, i<j<s, i,j,s=1,2,3,4,
\end{aligned}
\end{eqnarray*}
along with the differential consequences with respect to $x_i,i=1,2,3,4.$\\
\textbf{Type-4 invariant subspace} $ {^{4}_{4}\mathcal{\hat{W}}}_{l} $: Finally, the Type-4 linear space for the given operator $\mathcal{\hat{N}}_2[u]$ is defined in the form 
\begin{eqnarray}\begin{aligned}\label{4dt4}
{^{4}_{4}\mathcal{\hat{W}}}_{l}=&\text{Span}({_{4}\mathcal{\hat{S}}}^{4})
=\left\{\xi(x_1,x_2,x_3,x_4)= \lambda+\sum_{i=1}^{4}\lambda_{q_i}  \omega_{q_i}^{i}(x_i)\big{|} \lambda,\lambda_{q_{i}}\in \mathbb{R},\mathcal{D}_{x_i}[\xi]=0,
\right.\\&\left.
a_{i0}=0,
\dfrac{\partial^2\xi}{\partial x_{i}\partial x_{j}}=0,i<j,
q_{i}=1,\dots,n_{i}-1,i,j=1,2,3,4\right.\Biggr\},
\end{aligned}
\end{eqnarray}
 where the basis set
$ {_{4}\mathcal{\hat{S}}}^{4}=\left\{1,\omega_{1}^1(x_1),\dots,\omega_{n_{1}-1}^1(x_1),\dots,\omega_{n_{4}-1}^4(x_4)\right\}
$
and the dimension of the linear space ${^{4}_{4}\mathcal{\hat{W}}}_{l}$ is $\text{dim}({^{4}_{4}\mathcal{\hat{W}}}_{l})=n_1+n_{2}+n_3+n_4-3.$\\
The Type-4 linear space ${^{4}_{4}\mathcal{\hat{W}}}_{l}$ is said to be invariant under the given  operator $ \mathcal{\hat{N}}_2[u]$ if 
$\mathcal{\hat{N}}_2[u]\in {^{4}_{4}\mathcal{\hat{W}}}_{l}$ whenever $ u\in {^{4}_{4}\mathcal{\hat{W}}}_{l}.$  Hence the invariant conditions for the Type-4 linear space ${^{4}_{4}\mathcal{\hat{W}}}_{l}$ read as
\begin{eqnarray}\begin{aligned}\label{invc4dt4}
&\mathcal{D}_{x_i}\left[\mathcal{\hat{N}}_2[u]\right]=0,
\dfrac{\partial^2\mathcal{\hat{N}}_2[u]}{\partial x_{i}\partial x_{j}}=0,a_{i0}=0\,i<j,i,j=1,2,3,4
\\\text{whenever} \,
&\mathcal{D}_{x_i}[u]=0,a_{i0}=0,
\dfrac{\partial^2u}{\partial x_{i}\partial x_{j}}=0,i<j,i,j=1,2,3,4,
\end{aligned}
\end{eqnarray}
 along with the differential consequences with respect to $x_i,i=1,2,3,4.$
 \begin{nt}
It is interesting to note that
if $1 \in {_4^{1}\mathcal{\hat{W}}}_{l}$, then we obtain the following inclusion relation
\[ ^{4}_{4}\mathcal{\hat{W}}_l\subseteq{^{3}_{4}\mathcal{\hat{W}}}_{l}\subseteq {^{2}_{4}\mathcal{\hat{W}}}_{l} \subseteq{^{1}_{4}\mathcal{\hat{W}}}_{l}.
\]
 From this, we observe that ${_4^{j} \mathcal{\hat{W}}}_l $ is invariant with respect to the given differential operator $\mathcal{\hat{N}}_2[u]$, but the linear space $ {_4^{i} \mathcal{\hat{W}}}_l $ need not be an invariant under $\mathcal{\hat{N}}_2[u]$, where  $i<j,i,j=1,2,3,4$.\end{nt}
Next, we develop the generalization of the invariant subspace method for the generalized $(m+1)$-dimensional non-linear TFPDEs. The following section gives a systematic analysis to find the various types of linear spaces to the generalized $(m+1)$-dimensional non-linear TFPDEs.
\section{Generalized $(m+1)$-dimensional non-linear TFPDEs : Invariant subspace method }\label{sec4}
In this section, we consider the following sufficiently smooth generalized non-linear partial differential operator of order $k$, involving $m$-independent variables as follows:
\begin{eqnarray}\begin{aligned}\label{mdop}
\mathcal{\hat{N}}_3[u]\equiv&\mathcal{\hat{N}}\left( x_{1},x_{2},\ldots,x_{m},u,
\frac{\partial u}{\partial x_1},\frac{\partial u}{\partial x_2},\ldots,\frac{\partial^{k}u}{\partial x_1^{k_1}\partial x_2^{k_2}\cdots{\partial x_m^{k_m}}}\right),\end{aligned}
\end{eqnarray}
where $t>0$, $x_i\in\mathbb{R},i=1,2,\ldots,m$, $u=u(x_1,x_2,\cdots,x_m,t)$ and $\sum\limits_{i=1}^{m}k_i=k$, $k_i \in \mathbb{N},\ i=1,2,\ldots,m$.
Now, we define the $n$-dimensional linear space spanned by the $n$-linearly independent functions $ \xi_q(x_1,x_2,\ldots,x_m), q=1,2,\dots,n $, that is
\begin{eqnarray} \label{mdls}
\begin{aligned}
{_m\mathcal{\hat{W}}_n}&=\text{Span}(\mathcal{B})=\left\{\sum_{q=1}^{n}\lambda_q\xi_q(x_1,x_2,\ldots,x_m):\lambda_q \in \mathbb{R}, q = 1,2,\dots,n\right\},
\end{aligned}
\end{eqnarray}
where the basis set $\mathcal{B}= \{\xi_q(x_1,x_2,\ldots,x_m): q=1,2,\dots,n \}$ and $n$ is the dimension of the linear space ${_m\mathcal{\hat{W}}_n}$.\\
 The non-linear partial differential operator  $\mathcal{\hat{N}}_3[u]$ given in \eqref{mdop} preserves the linear space ${_m\mathcal{\hat{W}}_n}$ if $\mathcal{\hat{N}}_3[u] \in{_m\mathcal{\hat{W}}_n}$ whenever $ u\in {_m\mathcal{\hat{W}}_n}. $ If the linear space ${_m\mathcal{\hat{W}}_n}$ is invariant under the given differential operator \eqref{mdop}, then we have
 \[\mathcal{\hat{N}}_3\left[\sum\limits_{q=1}^{n}\lambda_q\xi_q(x_1,x_2,\ldots,x_m) \right]=\sum \limits^{n} _{q=1}\Lambda_q(\lambda_1,\lambda_2,\ldots,\lambda_n)\xi_q(x_1,x_2,\ldots,x_m)\in{_m\mathcal{\hat{W}}_n}, \]
where the functions $\lambda_q\in \mathbb{R}$ and the functions $\Lambda_q(\lambda_1,\lambda_2,\ldots,\lambda_n),q=1,2,\dots,n, $ denote the coefficients of expansion with respect to the basis $ \mathcal{B} .
$
\begin{thm}\label{thm3}
Let us consider the generalized $(m+1)$-dimensional non-linear TFPDE
\begin{equation}\label{mdpde}
\frac{\partial^{\alpha}u}{\partial t^{\alpha}}=\mathcal{\hat{N}}_3[u],\ \alpha>0,\ t>0,
\end{equation}
where $u={u}(x_1,x_2,\ldots,x_m,t)$. 
If the operator $\mathcal{\hat{N}}_3[u]$ preserves the space ${_m\mathcal{\hat{W}}_n}$ given in \eqref{mdls}, then there exists an exact solution to the equation \eqref{mdpde} in the finite separable form
\begin{equation}\label{mdsol}
{u}(x_1,x_2,\ldots,x_m,t)=\sum_{q=1}^{n}\Psi_q(t)\xi_q(x_1,x_2,\ldots,x_m),
\end{equation}
where the functions $\Psi_q(t),q=1,2,\dots,n$ satisfy the system of arbitrary-order ODEs
\begin{equation}\label{md4}
\frac{d^{\alpha}\Psi_q(t)}{dt^\alpha}=\Lambda_q(\Psi_1(t),\Psi_2(t),...,\Psi_n(t)),q=1,2,\dots,n.
\end{equation}.
\textbf{Proof:} Similar to the proof of theorem \ref{thm1}.
\end{thm}
Next, we explain how to define $m$-different types of linear spaces for the operator $\mathcal{\hat{N}}_3[u]$ given in \eqref{mdop}.
\subsection{Invariant subspaces admitted by the given differential operator \eqref{mdop}}
Here, let us first consider the homogeneous linear ODEs of order ${n_i}$ in the form
\begin{equation}\label{mdde}
 \mathcal{D}_{x_i}[\omega]\equiv \dfrac{d^{n_i}\omega}{d x_{i}^{n_i}}+ a_{in_i-1}\dfrac{d^{n_i-1}\omega}{d x_{i}^{n_i-1}}+\cdots+a_{i0}\omega=0,i=1,2,\dots,m
\end{equation}
with fundamental set of solutions as
$ _m\mathcal{S}_{x_i}=\{\omega_{q_{i}}^i(x_i): \mathcal{D}_{x_i}[\omega_{q_{i}}^i(x_i)]=0,q_i=1,2,\ldots,n_{i}\}$ and $ a_{ik} \in \mathbb{R}, \,k=0,1,\ldots,n_i-1,\, i=1,2,\ldots,m .$
Thus, we define the solution space of the $n_i$-th order homogeneous linear ODEs given in \eqref{mdde} as
\begin{eqnarray}\begin{aligned}\label{mdss}
	\mathcal{\hat{W}}^{x_i}_{n_i}=&\text{Span}(_m\mathcal{S}_{x_i})
=\left\{\sum_{q_i=1}^{n_i}\kappa_{q_i}\omega_{q_{i}}^i(x_i)|\mathcal{D}_{x_i}[\omega_{q_{i}}^i(x_i)]=0,\kappa_{q_i}\in \mathbb{R},q_i=1,2,\ldots,n_{i}
\right\}.
\end{aligned}
\end{eqnarray}
Now, we can define the $m$-different types of linear spaces for the non-linear partial differential operator $\mathcal{\hat{N}}_3[u]$ given in \eqref{mdop}.\\
\textbf{\textit{Type-1 invariant subspace} $ {_m^{1}\mathcal{\hat{W}}}_{l} $}:
First, let us define the Type-1 linear space
\begin{eqnarray}\begin{aligned}\label{mdt1}
{_m^{1}\mathcal{\hat{W}}}_{l}=&\text{Span}({_m\mathcal{\hat{S}}}^1)
=\left\{\xi(x_1,x_2,\dots,x_m)=\sum_{q_{1},\dots,q_m}\lambda_{q_{1},q_{2},\dots, q_{m}}\prod_{i=1}^{m}\omega_{q_{i}}^{i}(x_{i})  \big{|}
\right.\\&\left.
\mathcal{D}_{x_i}[\xi(x_1,x_2,\dots,x_m)]=0,
\lambda_{q_{1},q_{2},\dots, q_{m}}\in \mathbb{R},q_{i}=1,2,\dots,n_{i},i=1,\dots,m\right.\Biggr\},
\end{aligned}
\end{eqnarray}
where $l$ denotes the dimension of the linear space ${_m^{1}\mathcal{\hat{W}}}_{l}$, that is, $\text{dim}(_m^{1}\mathcal{\hat{W}}_l)=l=\prod\limits_{i=1}^{m} n_i$ and  the basis  \\
$
{_m\mathcal{\hat{S}}}^1=\left\{\prod\limits_{i=1}^{m}  \omega_{q_i}^{i}(x_i)\big{|}  \omega_{q_i}^{i}(x_i)\in {_m\mathcal{\hat{S}}_{x_i}},q_i=1,2,\ldots,n_i,i=1,2,\ldots,m\right\}.
$ \\
The Type-1 linear space ${_m^{1}\mathcal{\hat{W}}}_{l}$ is invariant under $ \mathcal{\hat{N}}_3[u]$ if $ \mathcal{\hat{N}}_3[u]\in {_m^{1} \mathcal{\hat{W}}}_l $ whenever $ u\in {_m^{1} \mathcal{\hat{W}}}_l .$ For this case, the obtained invariant conditions are as follows :
 \begin{eqnarray}\begin{aligned}
& \mathcal{D}_{x_i}\left[\mathcal{\hat{N}}_3[u]\right]=0,\,\, \text{whenever}\;
& { \mathcal{D}}_{x_i}[u]=0,i=1,2,\ldots,m,
\end{aligned} \end{eqnarray}
 along with the differential consequences with respect to $x_i,i=1,2,\dots,m.
 $
 \\
  \textbf{\textit{Type-2 invariant subspace} ${_m^{2}\mathcal{\hat{W}}}_{l}$}:
Now, let $\omega_{n_{r}}^r(x_{r})=1\in {_m\mathcal{{S}}_{x_{r}}},r=1,2,\dots,m$. Thus, we can define the Type-2 linear space as follows:
\begin{eqnarray}\begin{aligned}\label{mdt2}
{_m^{2}\mathcal{\hat{W}}}_{l}&=
\text{Span}({_m\mathcal{\hat{S}}}^2)
=\left\{\xi(x_1,x_2,\dots,x_m)
=\lambda+
\sum_{i=1}^{m}\sum_{q_{i}=1}^{n_i-1}\lambda_{q_{i}}^{i}\omega_{q_{i}}^{i}(x_{i})
+\ldots\right.
\\
 &\left.
+\sum_{\mathclap{\substack{i_j=1\\i_1<\dots<i_{m-1}}}}^m\left[\sum_{\mathrlap{q_{i_1},\dots,q_{i_{m-1}}}}\lambda_{q_{i_1},q_{i_2},\dots, q_{i_{m-1}}}^{i_{1},i_{2},\dots ,i_{m-1}}
\prod_{j=1}^{m-1}\omega_{q_{i_j}}^{i_j}(x_{i_j})  \right]
\big{|}
 \mathcal{D}_{x_{i_j}}[\xi]=0,
 \dfrac{\partial^{m}\xi}{\partial x_{1}\partial x_{2}\dots\partial x_{{m}}}=0,
 \right.\\
&\left.
a_{i_j0}=0,
\lambda_{q_{i_1},q_{i_2},
\dots ,q_{i_{m-1}}}^{i_{1},i_{2},\dots, i_{m-1}},\dots
\lambda_{q_{i}}^{i},\lambda\in \mathbb{R},
q_{i_j}=1,\dots,n_{i_j},i_j=1,\dots,m,j=1,\dots,m-1 \right.\Biggr\},
\end{aligned}
\end{eqnarray}
where the dimension of the linear space  $ {_m^{2}\mathcal{\hat{W}}}_{l} $ is $l$, that is,
\[\text{dim}({_m^{2}\mathcal{\hat{W}}}_{l})=l=1+\sum\limits _{i =1}^{m}(n_i-1) +\sum\limits _{\mathclap{\substack{i_r<i_s\\r,s=1,\dots,m-1\\i_r,i_s =1,2,\dots,m }}}(n_{i_r}-1)(n_{i_s}-1)+\dots+\sum\limits _{\mathclap{\substack{i_r=1\\r=1,\dots,m-1\\i_1<\dots<i_{m-1}}}}^m\left[\prod\limits _{j=1}^{m-1}(n_{i_j}-1)\right]
\]
and the basis set
$
 {_m\mathcal{\hat{S}}}^2=\bigcup\limits_{\mathclap{\substack{i_j=1\\i_1<\dots<i_{m-1}}}}^m\left
\{\prod\limits_{j=1}^{m-1}\omega_{q_{i_j}}^{i_j}(x_{i_{j}})\big{|}\omega_{q_{i_j}}^{i_j}(x_{i_{j}})\in {_m\mathcal{S}_{x_{i_j}}},
q_{i_{j}}=1,\ldots,n_{i_{j}},j=1,\dots,m-1\right\}.
$
For this case, the invariant conditions for the operator $\mathcal{\hat{N}}_3[u]$ given in \eqref{mdop} are obtained as follows:
\begin{eqnarray}\begin{aligned}
&\mathcal{D}_{x_{i}}\left[\mathcal{\hat{N}}_3[u]\right]=0,
\dfrac{\partial^{m} \mathcal{\hat{N}}_3[u] }{\partial x_{1}\partial x_{2}\dots\partial x_{{m}}}=0, a_{i0}=0,
\,\text{whenever}\\&
\mathcal{D}_{x_{i}}[u]=0,
\dfrac{\partial^{m} u }{\partial x_{1}\partial x_{2}\dots\partial x_{{m}}}=0, a_{i0}=0, i=1,2,\dots,m,
\end{aligned}
\end{eqnarray}
 along with the differential consequences with respect to $x_i,i=1,2,\dots,m.
 $
 \\
 \textbf{\textit{Type-3 invariant subspace} ${_m^{3}\mathcal{\hat{W}}}_{l}$}:
Let $\omega_{n_{r}}^r(x_{r})=1\in {_m\mathcal{{S}}_{x_{r}}},r=1,2,\dots,m$. Thus, we can construct the Type-3 invariant subspaces for the given operator \eqref{mdop} as
\begin{eqnarray}\begin{aligned}\label{mdt3}
{_m^{3}\mathcal{\hat{W}}}_{l}&=
\text{Span}({_m\mathcal{\hat{S}}}^3)
=\left\{\xi(x_1,x_2,\dots,x_m)=\lambda+\sum_{i=1}^{m}\sum_{q_{i}=1}^{n_i-1}\lambda_{q_{i}}^{i}\omega_{q_{i}}^{i}(x_{i})+\ldots\right.
\\
&\left.\begin{small}
 +\sum_{\mathclap{\substack{i_j=1\\i_1<\dots<i_{m-2}}}}^m\left[\sum_{\mathrlap{q_{i_1},\dots,q_{i_{m-2}}}}\lambda_{q_{i_1},q_{i_2},\dots, q_{i_{m-2}}}^{i_{1},i_{2},\dots ,i_{m-2}}\prod_{j=1}^{m-2}\omega_{q_{i_j}}^{i_j}(x_{i_j})  \right]\end{small}
 \big{|}\mathcal{D}_{x_{i_j}}[\xi]=0,
 \dfrac{\partial^{m-1}\xi}{\partial x_{i_1}\partial x_{i_2}\dots\partial x_{{i_{m-1}}}}=0,
  \right.\\
&\left.
a_{i_j0}=0,
 \lambda_{q_{i_1},q_{i_2},\dots ,q_{i_{m-2}}}^{i_{1},i_{2},\dots, i_{m-2}},\dots,\lambda_{q_{i}}^{i},\lambda\in \mathbb{R},
  q_{i_j}=1,\dots,n_{i_j},i_j=1,\dots,m,j=1,\dots,m-2\right.\Biggr\},
\end{aligned}
\end{eqnarray}
where the dimension of the linear space  ${ _m^{3}\mathcal{\hat{W}}}_{l} $ is given by
\[\text{dim}({_m^{3}\mathcal{\hat{W}}}_{l})=l=1+\sum\limits _{i =1}^{m}(n_i-1) +\sum\limits _{\mathclap{\substack{i_r<i_s\\r,s=1,\dots,m-2\\i_r,i_s =1,2,\dots,m }}}(n_{i_r}-1)(n_{i_s}-1)+\dots+\sum\limits _{\mathclap{\substack{i_r=1\\r=1,\dots,m-1\\i_1<\dots<i_{m-2}}}}^m\left[\prod\limits _{j=1}^{m-2}(n_{i_j}-1)\right]
\]
and the basis set
$
 {_m\mathcal{\hat{S}}}^3=\bigcup\limits_{\mathclap{\substack{i_j=1\\i_1<\dots<i_{m-2}}}}^m\left
\{\prod\limits_{j=1}^{m-2}\omega_{q_{i_j}}^{i_j}(x_{i_{j}})\big{|}\omega_{q_{i_j}}^{i_j}(x_{i_{j}})\in {_m\mathcal{S}_{x_{i_j}}},
q_{i_{j}}=1,2,\ldots,n_{i_{j}},j=1,\dots,m-2\right\}
$.\\
Thus, the invariant conditions for Type-3 linear space for given operator \eqref{mdop} read as follows:
\begin{eqnarray}\begin{aligned}
&\mathcal{D}_{x_{i_j}}\left[\mathcal{\hat{N}}_3[u]\right]=0,a_{i_j0}=0,
\dfrac{\partial^{m-1} \mathcal{\hat{N}}_3[u] }{\partial x_{i_1}\partial x_{i_2}\dots\partial x_{i_{m-1}}}=0,
\text{whenever}\, \\&
\mathcal{D}_{x_{i_j}}[u]=0,a_{i_j0}=0,
\dfrac{\partial^{m-1} u }{\partial x_{i_1}\partial x_{i_2}\dots\partial x_{i_{m-1}}}=0, i_j=1,2,\dots,m,j=1,2,\dots,m-1
\end{aligned}
\end{eqnarray}
 along with the differential consequences with respect to
  $x_i,i=1,2,\dots,m.$
\\
Now, we can construct all possible types of linear spaces for the given non-linear partial differential operator \eqref{mdop} in a single form except Type-1 linear space that is given below.
\\
\textbf{\textit{Type-$(p+1)$ invariant subspace} ${_m^{p+1}\mathcal{\hat{W}}}_{l} $ ($p=1,2,\dots,m-1$)}:
Now, let $\omega_{n_{r}}^r(x_{r})=1\in {_m\mathcal{{S}}_{x_{r}}},r=1,2,\dots,m$. Thus, we can construct the Type-$(p+1)$ invariant subspaces for the operator $\mathcal{\hat{N}}_3[u]$ given in \eqref{mdop} as
\begin{eqnarray}\begin{aligned}\label{mdtp+1}
{_m^{p+1}\mathcal{\hat{W}}}_{l}&
=\text{Span}({_m\mathcal{\hat{S}}}^{p+1})
=\left\{\xi(x_1,x_2,\dots,x_m)=\lambda+\sum_{i=1}^{m}\sum_{q_{i}=1}^{n_i-1}\lambda_{q_{i}}^{i}\omega_{q_{i}}^{i}(x_{i})+\ldots\right.
\\
&\left. \begin{small}
 +\sum_{\mathclap{\substack{i_j=1\\j=1,\dots,m-p\\i_1<\dots<i_{m-p}}}}^m\left[\sum_{\mathrlap{q_{i_1},\dots,q_{i_{m-p}}}}\lambda_{q_{i_1},q_{i_2},\dots, q_{i_{m-p}}}^{i_{1},i_{2},\dots ,i_{m-p}}\prod_{j=1}^{m-p}\omega_{q_{i_j}}^{i_j}(x_{i_j})  \right]\end{small}
 \big{|}
 \mathcal{D}_{x_{i_j}}[\xi]=0,\dfrac{\partial^{m-p+1}\xi }{\partial x_{i_1}\partial x_{i_2}\dots\partial x_{i_{m-p+1}}}=0,
 \right.\\
&\left.
a_{i_j0}=0,
\lambda_{q_{i_1},q_{i_2},\dots ,q_{i_{m-p}}}^{i_{1},i_{2},\dots, i_{m-p}},\dots,\lambda\in \mathbb{R},
q_{i_j}=1,\dots,n_{i_j},i_j=1,\dots,m,j=1,\dots,m-p\right.\Biggr\},
\end{aligned}
\end{eqnarray}
where the dimension of linear space ${_m^{p+1}\mathcal{\hat{W}}}_{l} $ is given by
\[\text{dim}({_m^{p+1}\mathcal{\hat{W}}}_{l})=l=1+\sum\limits _{i =1}^{m}(n_i-1) +\sum\limits _{\mathclap{\substack{i_r<i_s\\r,s=1,\dots,m-p\\i_r,i_s =1,2,\dots,m }}}(n_{i_r}-1)(n_{i_s}-1)+\dots+\sum\limits _{\mathclap{\substack{i_r=1\\r=1,\dots,m-p\\i_1<\dots<i_{m-p}}}}^m\left[\prod\limits _{j=1}^{m-p}(n_{i_j}-1)\right]
\]
and the basis set
$ {_m\mathcal{\hat{S}}}^{p+1}=\bigcup\limits_{\mathclap{\substack{i_j=1\\j=1,\dots,m-p\\i_1<\dots<i_{m-p}}}}^m\left
\{\prod\limits_{j=1}^{m-p}\omega_{q_{i_j}}^{i_j}(x_{i_{j}})\big{|}\omega_{q_{i_j}}^{i_j}(x_{i_{j}})\in {_m\mathcal{\hat{S}}_{x_{i_j}}},
q_{i_{j}}=1,\ldots,n_{i_{j}}\right\},$ for each $p=1,2,\dots,m-1.$
\\
 The Type-$(p+1)$  linear space ${_m^{p+1}\mathcal{\hat{W}}}_{l} $ is invariant under the given differential operator $\mathcal{\hat{N}}_3[u]$ if $ \mathcal{\hat{N}}_3[u]\in {_m^{p+1} \mathcal{\hat{W}}}$ whenever $u \in {_m ^{p+1} \mathcal{\hat{W}}}_l ,$ for $p=1,2,\dots,m-1$. Thus, the invariant conditions read as
\begin{eqnarray}\begin{aligned}
&\mathcal{D}_{x_{i_j}}\left[\mathcal{\hat{N}}_3[u]\right]=0,
\dfrac{\partial^{m-p+1} \mathcal{\hat{N}}_3[u] }{\partial x_{i_1}\partial x_{i_2}\dots\partial x_{i_{m-p+1}}}=0, a_{i_j0}=0,
\,\text{whenever}\\
&\mathcal{D}_{x_{i_j}}[u]=0,
\dfrac{\partial^{m-p+1} u }{\partial x_{i_1}\partial x_{i_2}\dots\partial x_{i_{m-p+1}}}=0,a_{i_j0}=0,i_j=1,\dots,m,j=1,2,\dots,m-p+1,
\end{aligned}
\end{eqnarray}
 where $p=1,2,\dots,m-1$ along with the differential consequences with respect to
$x_i,i=1,2,\dots,m.$

\begin{nt}  It is important to note that if $1\in{_m^{1} \mathcal{\hat{W}}_l}$, then we obtain the following inclusion relation
 \[  {_m^{m} \mathcal{\hat{W}}_l} \subseteq {_m^{m-1} \mathcal{\hat{W}}_l}\subseteq\ldots\subseteq{_m^{2} \mathcal{\hat{W}}_l}\subseteq{_m^{1} \mathcal{\hat{W}}_l}.
\]
 From this, we can observe that the linear space ${_m^{j} \mathcal{\hat{W}}_l} $ is invariant under the given differential operator $\mathcal{\hat{N}}_3[u]$, but the linear space ${_m^{i} \mathcal{\hat{W}}_l}$ need not be an invariant under $\mathcal{\hat{N}}_3[u]$ where $i<j,$ $\;i,j=1,2,\dots,m$.\end{nt}
Next, we explain how to find various types of linear spaces with different dimensions for the well-known generalized $(3+1)$-dimensional non-linear time-fractional DCW equation \eqref{cdeqn}.
\section{Determining the invariant subspaces for the given non-linear time-fractional DCW equation \eqref{cdeqn} }
This section gives a detailed algorithmic approach to find the various types of invariant subspaces for the generalized $(3+1)$-dimensional non-linear time-fractional DCW equation \eqref{cdeqn}.
Hence the obtained non-linear differential operator is in the form
\begin{equation}\label{cdop}
\mathcal{\hat{H}}[u]=
 \sum\limits^{3}_{r=1}\dfrac{\partial  }{\partial x_r}\left( P_r(u)\frac{\partial u}{\partial x_r}\right)+\sum\limits^{3}_{r=1}Q_r(u)\left(\frac{\partial u}{\partial x_r}\right).
\end{equation}
For this work, let us consider the following values of $(n_1,n_2,n_3):$
 \begin{equation}\label{n}
  \left\{(1,1,1),(2,2,2),(2,2,3),(2,3,2),(3,2,2),(2,3,3),(3,2,3),(3,3,2),(3,3,3)\right\}.
 \end{equation}
Let us first consider the case $n_i=2$, $i=1,2,3$.
 Next, we wish to explain how to construct the three types of linear spaces for the given non-linear partial differential operator $\mathcal{\hat{H}}[u]$ in an algorithmic way.\\
Let us define solution spaces of the linear homogeneous ODEs of order $n_i=2$ $(i=1,2,3)$ as
$$\mathcal{\hat{W}}_{n_i}^{x_i}=\text{Span}({_3\mathcal{S}}_{x_i}),$$
where ${_3S}_{x_i}=\{\omega_{1}^i(x_i),\omega_{2}^i(x_i)\}$ is the fundamental solution set of the following homogeneous ODEs of order $n_i=2$ $(i=1,2,3)$
\begin{eqnarray}\label{cdode}
\mathcal{D}_{x_1}[\omega]\equiv \dfrac{d^{2}\omega}{d x_{1}^{2}}+ a_{11}\dfrac{d\omega}{d x_{1}}+a_{10}\omega=0,\\
\mathcal{D}_{x_2}[\omega]\equiv \dfrac{d^{2}\omega}{d x_{2}^{2}}+ a_{21}\dfrac{d\omega}{d x_{2}}+a_{20}\omega=0,\\
\mathcal{D}_{x_3}[\omega]\equiv \dfrac{d^{2}\omega}{d x_{3}^{2}}+ a_{31}\dfrac{d\omega}{d x_{3}}+a_{30}\omega=0
\end{eqnarray}
 and $ a_{ij} \in \mathbb{R},$ $j=0,1,\, i=1,2,3.$\\
 Now, we explain how to define three types of linear spaces for the given non-linear partial differential operator \eqref{cdop} using the obtained solution spaces $\mathcal{\hat{W}}_{n_i}^{x_i}$ $(i=1,2,3)$.\\
\textbf{Type-1 linear space and their invariant conditions}:
The Type-1 linear space is defined as follows
\begin{eqnarray}
\begin{aligned}\label{type1}
^1_3\mathcal{\hat{W}}_l=&\text{Span}\left\{  \omega_{1}^{1}(x_1)  \omega_{1}^{2}(x_2)  \omega_{1}^{3}(x_3), \omega_{2}^{1}(x_1)  \omega_{1}^{2}(x_2)  \omega_{1}^{3}(x_3), \omega_{1}^{1}(x_1) \omega_{2}^{2}(x_2)  \omega_{1}^{3}(x_3),
\right.
\\&
\omega_{1}^{1}(x_1)  \omega_{2}^{2}(x_2) \omega_{2}^{3}(x_3),
  \omega_{1}^{1}(x_1)  \omega_{1}^{2}(x_2)  \omega_{2}^{3}(x_3), \omega_{2}^{1}(x_1)  \omega_{2}^{2}(x_2)  \omega_{1}^{3}(x_3),\\&
\left.  \omega_{2}^{1}(x_1) \omega_{1}^{2}(x_2) \omega_{2}^{3}(x_3), \omega_{2}^{1}(x_1) \omega_{2}^{2}(x_2) \omega_{2}^{3}(x_3) \right\}.
\end{aligned}
\end{eqnarray}
It should be noted that the operator $\mathcal{\hat{H}}[u]$ given in \eqref{cdop} preserves the linear space $^1_3\mathcal{\hat{W}}_l $ given in \eqref{type1} if it satisfies the following conditions:
\begin{eqnarray}
\mathcal{D}_{x_i}[ \mathcal{\hat{H}}[u]]=0
\; \text{whenever}\;
\mathcal{D}_{x_i}[u]=0,i=1,2,3.
\end{eqnarray}
Additionally, we observe that the dimension of the Type-1 invariant subspace is $8$, that is $\text{dim}(^1_3\mathcal{\hat{W}}_l)=8$.\\
\textbf{Type-2 linear space and their invariant conditions}:
The Type-2 linear space is defined in the following form
\begin{eqnarray}
\begin{aligned}\label{type2}
^2_3\mathcal{\hat{W}}_l=&\text{Span}\left\{1,  \omega_{1}^{1}(x_1) , \omega_{1}^{2}(x_2),  \omega_{1}^{3}(x_3), \omega_{1}^{1}(x_1)  \omega_{1}^{2}(x_2) , \omega_{1}^{1}(x_1)  \omega_{1}^{3}(x_3),
  \omega_{1}^{2}(x_2)  \omega_{1}^{3}(x_3)\right\}.
\end{aligned}
\end{eqnarray}
Thus, the non-linear partial differential operator $\mathcal{\hat{H}}[u]$ given in \eqref{cdop} preserves the linear space $^2_3\mathcal{\hat{W}}_l $ given in \eqref{type2} if the following conditions hold,
\begin{eqnarray}
\begin{aligned}
\mathcal{D}_{x_i}[ \mathcal{\hat{H}}[u]]=0, a_{i0}=0,i=1,2,3 \; \text{and}\,
\dfrac{\partial^3 \mathcal{\hat{H}}[u]}{\partial x_1 \partial x_2 \partial x_3} =0\\
\text{whenever}\;
\mathcal{D}_{x_i}[u]=0,a_{i0}=0,i=1,2,3 \;\text{and}\,
\dfrac{\partial ^3 u}{\partial x_1 \partial x_2 \partial x_3} =0.
\end{aligned}
\end{eqnarray}
In addition, we find that $\text{dim}(^2_3\mathcal{\hat{W}}_l)=7$.\\
\textbf{Type-3 linear space and their invariant conditions}:
Finally, we define the Type-3 linear space
\begin{eqnarray}
\begin{aligned}\label{type3}
^3\mathcal{\hat{W}}_l=&\text{Span}\left\{1,  \omega_{1}^{1}(x_1) , \omega_{1}^{2}(x_2),  \omega_{1}^{3}(x_3) \right\}.
\end{aligned}
\end{eqnarray}
The given differential operator $\mathcal{\hat{H}}[u]$ preserves the Type-3 linear space $^3_3\mathcal{\hat{W}}_l $  given in \eqref{type3} if it satisfies the following invariant conditions:
\begin{eqnarray}
\begin{aligned}\label{cdt3}
\mathcal{D}_{x_i}[ \mathcal{\hat{H}}[u]]=0, a_{i0}=0,i=1,2,3,
\dfrac{\partial^2 \mathcal{\hat{H}}[u]}{\partial x_1 \partial x_2}=0,\dfrac{\partial^2 \mathcal{\hat{H}}[u]}{\partial x_1 \partial x_3}=0\,\text{and}\,\,\dfrac{\partial^2 \mathcal{\hat{H}}[u]}{\partial x_3 \partial x_2}=0\\
\text{whenever}\;
\mathcal{D}_{x_i}[u]=0, a_{i0}=0,i=1,2,3,\,
\dfrac{\partial ^2 u}{\partial x_1 \partial x_2 } =0,
 \dfrac{\partial ^2 u}{\partial x_1 \partial x_3 } =0 \, \text{and}\,\dfrac{\partial ^2 u}{\partial x_3 \partial x_2 } =0.
\end{aligned}
\end{eqnarray}
Note that $\text{dim}(^3_3\mathcal{\hat{W}}_l)=4$.\\
Now, our aim is to find the Type-3 invariant subspaces for the operator $\mathcal{\hat{H}}[u]$ given in \eqref{cdop}. Thus, the Type-3 invariance criteria \eqref{cdt3} for the given operator \eqref{cdop} read in the following form,
\begin{eqnarray}
\begin{aligned}\label{d1g}
&D^{(3)}[P_1(u)]u_{x_1}^4+
\left[-5a_{11}D^{(2)}[P_1(u)]+D^{(2)}[Q_1(u)]\right]u_{x_1}^{3}+\left.\Biggl[4a_{11}^{2}D[P_1(u)]\right.
\\
-&\left. 2a_{11}D[Q_1(u)]+
\sum_{i=2,3}\left.\biggl(D^{(3)}[P_i(u)]u_{x_i}^2+
\left[-a_{i1}D^{(2)}[P_i(u)]+D^{(2)}[Q_i(u)]\right]u_{x_i}\right.\biggr)
\right]u_{x_1}^2
=0,
\end{aligned}
\end{eqnarray}
\begin{eqnarray}
\begin{aligned}\label{d2g}
&D^{(3)}[P_2(u)]u_{x_2}^4+
\left[-5a_{21}D^{(2)}[P_2(u)]+D^{(2)}[Q_2(u)]\right]u_{x_2}^{3}+
\left.\Biggl[4a_{21}^{2}D[P_2(u)]\right.
\\
-&\left.
2a_{21}D[Q_2(u)]+
\sum_{i=1,3}\left.\biggl(D^{(3)}[P_i(u)]u_{x_i}^2+
\left[-a_{i1}D^{(2)}[P_i(u)]+D^{(2)}[Q_i(u)]\right]u_{x_i}
\right.\biggr)\right]u_{x_2}^2=0,
\end{aligned}
\end{eqnarray}

\begin{eqnarray}
\begin{aligned}\label{d3g}
&D^{(3)}[P_3(u)]u_{x_3}^4+
\left[-5a_{31}D^{(2)}[P_3(u)]+D^{(2)}[Q_3(u)]\right]u_{x_3}^{3}+\left.\Biggl[4a_{31}^{2}D[P_3(u)]
\right.
\\
-&\left.
2a_{31}D[Q_3(u)]+
\sum_{i=1,2}\left.\biggl(D^{(3)}[P_i(u)]u_{x_i}^2+
\left[-a_{i1}D^{(2)}[P_i(u)]+D^{(2)}[Q_i(u)]\right]u_{x_i}\right.\biggr)
\right]u_{x_3}^2=0,
\end{aligned}
\end{eqnarray}
\begin{eqnarray}
\begin{aligned}\label{D4g}
&\Bigg[D^{(3)}[P_3(u)]u_{x_3}^2+
\left[-a_{31}D^{(2)}[P_3(u)]+D^{(2)}[Q_3(u)]\right]u_{x_3}\Bigg]u_{x_1 }u_{x_2}
+
\sum_{i,j=1,2}\Bigg[D^{(3)}[P_i(u)]u_{x_i}^{3}
\\
+&
\left[-3a_{i1}D^{(2)}[P_i(u)]+D^{(2)}[Q_i(u)]\right]u_{x_i}^{2}\Bigg]u_{x_j}
+\sum_{i=1,2}\left[a_{i1}^2D[P_i(u)]-a_{i1}D[Q_i(u)]\right]u_{x_1}u_{x_2 }=0,
\end{aligned}
\end{eqnarray}
\begin{eqnarray}
\begin{aligned}\label{D5g}
&\Bigg[D^{(3)}[P_2(u)]u_{x_2}^2+
\left[-a_{21}D^{(2)}[P_2(u)]+D^{(2)}[Q_2(u)]\right]u_{x_2}\Bigg]u_{x_3 }u_{x_1}
+
\sum_{i,j=1,3}\Bigg[D^{(3)}[P_i(u)]u_{x_i}^{3}
\\
+&
\left[-3a_{i1}D^{(2)}[P_i(u)]+D^{(2)}[Q_i(u)]\right]u_{x_i}^{2}\Bigg]u_{x_j}
+\sum_{i=1,3}\left[a_{i1}^2D[P_i(u)]-a_{i1}D[Q_i(u)]\right]u_{x_3}u_{x_1 }=0,
\end{aligned}
\end{eqnarray}
\begin{eqnarray}
\begin{aligned}\label{d6g}
&\Bigg[D^{(3)}[P_1(u)]u_{x_1}^2+
\left[-a_{11}D^{(2)}[P_1(u)]+D^{(2)}[Q_1(u)]\right]u_{x_1}\Bigg]u_{x_3 }u_{x_2}
+
\sum_{i,j=2,3}\Bigg[D^{(3)}[P_i(u)]u_{x_i}^{3}
\\
+&
\left[-3a_{i1}D^{(2)}[P_i(u)]+D^{(2)}[Q_i(u)]\right]u_{x_i}^{2}\Bigg]u_{x_j}
+\sum_{i=2,3}\left[a_{i1}^2D[P_i(u)]-a_{i1}D[Q_i(u)]\right]u_{x_3}u_{x_2 }=0,
\end{aligned}
\end{eqnarray}
where $u_{x_i}=\dfrac{\partial u}{\partial x_i},u_{x_ix_i}=\dfrac{\partial^2 u}{\partial x_i^2}$ and $ D^{(j)}[P_i(u)] ,D^{(j)}[Q_i(u)]$ denote the $j^{th}$ total derivative of the corresponding functions for $i=1,2,3$ and $j=1,2,3$. The above equations \eqref{d1g}-\eqref{d6g} give an over-determined system of ODEs. The obtained over-determined system of ODEs cannot be solved in general. Thus, we consider the differential operator \eqref{cdop} with cubic non-linearities, for simplicity.
Hence we obtain the functions $P_r(u)$ and $Q_r(u)$ $(r=1,2,3)$ in the forms,\\
\begin{tabular}{cc}
\label{cdopt}
$\left\{
\begin{array}{ll}
P_1(u)=\tau_{12}u^2+\tau_{11}u+\tau_{10}\\
P_2(u)=\tau_{22}u^2+\tau_{21}u+\tau_{20}\\
P_3(u)=\tau_{32}u^2+\tau_{31}u+\tau_{30}
\end{array}
\right.
$ &
$\left\{
\begin{array}{ll}
Q_1(u)=\rho_{12}u^2+\rho_{11}u+\rho_{10}\\
Q_2(u)=\rho_{22}u^2+\rho_{21}u+\rho_{20}\\
Q_3(u)=\rho_{32}u^2+\rho_{31}u+\rho_{30}\\
\end{array}
\right.
$ \\
\end{tabular}
\newline
where $\tau_{ij},\rho_{ij} \in \mathbb{R},i=1,2,3,j=0,1,2.$\\
Substituting the above functions $P_i(u) ,Q_i(u) ,i=1,2,3$ into the equations \eqref{d1g}-\eqref{d6g}, we get the following over-determined system of equations
\begin{eqnarray}
\begin{aligned}
&10a_{11}\tau_{12}-2\rho_{12}=0,
10a_{21}\tau_{22}-2\rho_{22}=0,
10a_{31}\tau_{32}-2\rho_{32}=0,
2a_{11}\tau_{12}-2\rho_{12}=0,
\\&
-2a_{21}\tau_{22}+2\rho_{22}=0,
-2a_{31}\tau_{32}+2\rho_{32}=0,
-6a_{11}\tau_{12}+\rho_{12}=0,
-6a_{21}\tau_{22}+\rho_{22}=0,\\
&
6a_{31}\tau_{32}-\rho_{32}=0,
8a_{11}^2\tau_{12}-4a_{11}\rho_{12}=0,
8a_{21}^2\tau_{22}-4a_{21}\rho_{22}=0,
8a_{31}^2\tau_{32}-4a_{31}\rho_{32}=0,\\
&
4a_{11}^2\tau_{11}-2a_{11}\rho_{11}=0,
4a_{21}^2\tau_{21}-2a_{21}\rho_{21}=0,
4a_{31}^2\tau_{31}-2a_{31}\rho_{31}=0,\\
&
a_{11}^{2}\tau_{11}+a_{21}^{2}\tau_{21}-a_{11}\rho_{11}-a_{21}\rho_{21}=0,
a_{11}^{2}\tau_{11}+a_{31}^{2}\tau_{31}-a_{31}\rho_{31}-a_{11}\rho_{11}=0,\\
&
a_{31}^{2}\tau_{31}+a_{21}^{2}\tau_{21}-a_{31}\rho_{31}-a_{21}\rho_{21}=0,
a_{11}^{2}\tau_{12}+a_{21}^{2}\tau_{22}-a_{11}\rho_{12}-a_{21}\rho_{22}=0,
\\
&
a_{11}^{2}\tau_{12}+a_{31}^{2}\tau_{32}-a_{31}\rho_{32}-a_{11}\rho_{12}=0,
a_{31}^{2}\tau_{32}+a_{21}^{2}\tau_{22}-a_{31}\rho_{32}-a_{21}\rho_{22}=0.
\end{aligned}
\end{eqnarray}
Solving the over-determined system of equations, we obtain the various types of linear spaces with their corresponding non-linear partial differential operators that are listed in Table \ref{c1}.
 Similarly, we can construct the various types of linear spaces for the given operator \eqref{cdop} with quadratic and cubic non-linearities for different values of $(n_1,n_2,n_3)$ given in \eqref{n} which are listed in the Tables \ref{q1}-\ref{c4}. 

\begin{sidewaystable}
\caption{Invariant subspaces and their cubic non-linear diffusion-convection-wave equation \eqref{cdeqn}}
\begin{center}
\begin{tabular}{cccc}
\toprule
   $P_r(u),r=1,2,3 $ &$ Q_r(u),r=1,2,3  $  & Invariant Subspaces \\
\midrule
1. $\left\{
\begin{array}{ll}
P_1(u)=\tau_{12}u^{2}+\tau_{11}u+\tau_{10}\\
P_2(u)=\tau_{22}u^{2}+\tau_{21}u+\tau_{20}\\
P_3(u)=\tau_{32}u^{2}+\tau_{31}u+\tau_{30}
\end{array}
\right.
$ &
$\left\{
\begin{array}{ll}
Q_1(u)=\rho_{11}u+\rho_{10}\\
Q_2(u)=\rho_{21}u+\rho_{20}\\
Q_3(u)=\rho_{31}u+\rho_{30}\\
\end{array}
\right.
$  & $

                                                                                                                                   \begin{array}{ll}
                                                                                                                                                                                                                                                                 {_3^3\mathcal{\hat{W}}_{4}}=\text{Span}\left\{1,x_{1},x_{2},x_{3}\right\}

                                                                                                                                   \end{array}
                                                                                                                                                                                                                                                           $\\
\bottomrule
\end{tabular}
\end{center}
\label{c1}
\end{sidewaystable}

\begin{sidewaystable}
\caption{Invariant subspaces and their corresponding quadratic non-linear diffusion-convection-wave equation \eqref{cdeqn}.}
\begin{center}
\begin{tabular}{cccc}
\toprule
   $P_r(u),r=1,2,3 $ &$ Q_r(u),r=1,2,3  $  & Invariant Subspaces \\
\midrule
1. $\left\{
\begin{array}{ll}
P_1(u)=\left(-\sum\limits_{r=2}^{3}\frac{a_{r0}^2\tau_{r1}}{a_{10}^2}
       \right.\\\left.
      \,+\sum\limits_{r=1}^{3}\frac{a_{r0}\rho_{r1}}{2a_{10}^2}\right)u+\tau_{10}\\
P_2(u)=\tau_{21}u+\tau_{20}\\
P_3(u)=\tau_{31}u+\tau_{30}
\end{array}
\right.
$ &
$\left\{
\begin{array}{ll}
Q_1(u)=\rho_{11}u+\rho_{10}\\
Q_2(u)=\rho_{21}u+\rho_{20}\\
Q_3(u)=\rho_{31}u+\rho_{30}\\
\end{array}
\right.
$  & $

                                                                                                                                   \begin{array}{ll}
                                                                                                                                                                                                                                                                 {_3^1\mathcal{\hat{W}}_{1}}=\text{Span}\left\{e^{-(a_{10}x_1+a_{20}x_2+a_{30}x_3)}\right\}

                                                                                                                                   \end{array}
                                                                                                                                                                                                                                                           $\\
\bottomrule
\end{tabular}
\end{center}
\label{q1}
\end{sidewaystable}

\begin{sidewaystable}
\caption{Invariant subspaces and their corresponding quadratic non-linear diffusion-convection-wave equation \eqref{cdeqn}}
\begin{center}
\begin{tabular}{cccc}
\toprule
   $P_r(u),r=1,2,3 $ &$ Q_r(u),r=1,2,3  $  & Invariant Subspaces \\
\midrule
8.$\left\{
\begin{array}{ll}
P_1(u)=\tau_{11}u+\tau_{10}\\
P_2(u)=\tau_{21}u+\tau_{20}\\
P_3(u)=\tau_{30}
\end{array}
\right.
$ &
$\left\{
\begin{array}{ll}
Q_1(u)=\rho_{10}\\
Q_2(u)=\rho_{21}u+\rho_{20}\\
Q_3(u)=\rho_{30}\\
\end{array}
\right.
$  & $
\left\{
                                                                                                                                   \begin{array}{ll}
                                                                                                                                                                                                                                                                {_3^3\mathcal{\hat{W}}_{5}}=\text{Span}\left\{1,x_{1},\frac{x_1^2}{2},x_{2},e^{-a_{31}x_{3}}\right\}
\\
 {_3^3\mathcal{\hat{W}}_{6}}=\text{Span}\left\{1,x_{1},\frac{x_1^2}{2},x_{2},sin(\sqrt{a_{31}}x_{3}),cos(\sqrt{a_{31}}x_{3})
\right\}

\\
 {_3^3\mathcal{\hat{W}}_{6}}=\text{Span}\left\{1,x_{1},\frac{x_1^2}{2},x_{2},e^{-\frac{1}{2}(a_{32}-\sqrt{a_{32}^{2}-4a_{31}})x_{3}},e^{-\frac{1}{2}(a_{32}+\sqrt{a_{32}^{2}-4a_{31}})x_{3}}\right\}

                                                                                                                                   \end{array}
                                                                                                                                   \right.
                                                                                                                                                                                                                                                           $\\
\bottomrule
\end{tabular}
\end{center}
\label{q2}
\end{sidewaystable}

\begin{sidewaystable}
\caption{Invariant subspaces and their corresponding quadratic non-linear diffusion-convection-wave equation \eqref{cdeqn}}
\begin{center}
\begin{tabular}{cccc}
\toprule
   $P_r(u),Q_r(u),r=1,2,3  $  & Invariant Subspaces \\
\bottomrule
\end{tabular}
\end{center}
\label{q3}
\end{sidewaystable}

\begin{sidewaystable}
\caption{Invariant subspaces and their corresponding quadratic non-linear diffusion-convection-wave equation \eqref{cdeqn}}
\begin{center}
\begin{tabular}{cccc}
\toprule
   $P_r(u),Q_r(u),r=1,2,3  $  & Invariant Subspaces \\
\bottomrule
\end{tabular}
\end{center}
\label{q4}
\end{sidewaystable}

\begin{sidewaystable}
\caption{Invariant subspaces and their corresponding quadratic non-linear diffusion-convection-wave equation \eqref{cdeqn}}
\begin{center}
\begin{tabular}{cccc}
\toprule
   $P_r(u),r=1,2,3 $ &$ Q_r(u),r=1,2,3  $  & Invariant Subspaces \\
\midrule

20.$\left\{
\begin{array}{ll}
P_1(u)=\tau_{11}u+\tau_{10}\\
P_2(u)=\tau_{21}u+\tau_{20}\\
P_3(u)=\tau_{31}u+\tau_{30}
\end{array}
\right.
$ &
$\left\{
\begin{array}{ll}
Q_1(u)=\rho_{11}u+\rho_{10}\\
Q_2(u)=\rho_{21}u+\rho_{20}\\
Q_3(u)=\rho_{31}u+\rho_{30}\\
\end{array}
\right.
$  & $

                                                                                                                                   \begin{array}{ll}
                                                                                                                                                                                                                                                                 {_3^3\mathcal{\hat{W}}_{4}}=\text{Span}\left\{1,x_{1},x_{2},x_{3}\right\}

                                                                                                                                   \end{array}
                                                                                                                                                                                                                                                           $\\
\bottomrule
\end{tabular}
\end{center}
\label{q5}
\end{sidewaystable}

\begin{sidewaystable}
\caption{Invariant subspaces and their corresponding cubic non-linear diffusion-convection-wave equation \eqref{cdeqn}}
\begin{center}
\begin{tabular}{cccc}
\toprule
   $P_r(u),r=1,2,3 $ &$ Q_r(u),r=1,2,3  $  & Invariant Subspaces \\
\midrule
1. $\left\{
\begin{array}{ll}
P_1(u)=\tau_{12}u^{2}+\tau_{11}u+\tau_{10}\\
P_2(u)=\tau_{22}u^{2}+\tau_{21}u+\tau_{20}\\
P_3(u)=\tau_{31}u+\tau_{30}
\end{array}
\right.
$ &
$\left\{
\begin{array}{ll}
Q_1(u)=\rho_{11}u+\rho_{10}\\
Q_2(u)=\rho_{21}u+\rho_{20}\\
Q_3(u)=\rho_{30}\\
\end{array}
\right.
$  & $

                                                                                                                                   \begin{array}{ll}
                                                                                                                                                                                                                                                                 {_3^3\mathcal{\hat{W}}_{5}}=\text{Span}\left\{1,x_{1},x_{2},x_{3},\frac{x_3^2}{2}\right\}

                                                                                                                                   \end{array}
                                                                                                                                                                                                                                                           $\\
\bottomrule
\end{tabular}
\end{center}
\label{c2}
\end{sidewaystable}

\begin{sidewaystable}
\caption{Invariant subspaces and their corresponding cubic non-linear diffusion-convection-wave equation \eqref{cdeqn}}
\begin{center}
\begin{tabular}{cccc}
\toprule
   $P_r(u),r=1,2,3 $ &$ Q_r(u),r=1,2,3  $  & Invariant Subspaces \\
\midrule
 9. $\left\{
\begin{array}{ll}
P_1(u)=\tau_{12}u^{2}+\tau_{11}u+\tau_{10}\\
P_2(u)=\tau_{22}u^{2}+\tau_{21}u+\tau_{20}\\
P_3(u)=\tau_{30}
\end{array}
\right.
$ &
$\left\{
\begin{array}{ll}
Q_1(u)=\rho_{11}u+\rho_{10}\\
Q_2(u)=\rho_{21}u+\rho_{20}\\
Q_3(u)=\rho_{30}\\
\end{array}
\right.
$  & $

\left\{
                                                                                                                                   \begin{array}{ll}
                                                                                                                                                                                                                                                                {_3^3\mathcal{\hat{W}}_{5}}=\text{Span}\left\{1,x_{1},x_{2},sin(\sqrt{a_{31}}x_{3}),cos(\sqrt{a_{31}}x_{3})\right\}
\\
 {_3^3\mathcal{\hat{W}}_{5}}=\text{Span}\left\{1,x_{1},x_{2},e^{-\frac{1}{2}(a_{32}-\sqrt{a_{32}^{2}-4a_{31}})x_{3}}, e^{-\frac{1}{2}(a_{32}+\sqrt{a_{32}^{2}-4a_{31}})x_{3}}\right\}

                                                                                                                                   \end{array}
                                                                                                                                                                                                                                                          \right. $\\
\bottomrule
\end{tabular}
\end{center}
\label{c3}
\end{sidewaystable}

\begin{sidewaystable}
\caption{Invariant subspaces and their corresponding cubic non-linear diffusion-convection-wave equation \eqref{cdop}}
\begin{center}
\begin{tabular}{cccc}
\toprule
   $P_r(u), $ $ Q_r(u),r=1,2,3  $  & Invariant Subspaces \\
\bottomrule
\end{tabular}
\end{center}
\label{c4}
\end{sidewaystable}
\section{Exact solutions of \eqref{cdeqn}}
This section explains how to find the exact solutions of the given equation \eqref{cdeqn} along with appropriate initial conditions using the invariant subspaces that are listed in the previous section. Here we explain how we can construct a variety of exact solutions for the underlying equation using the various types of invariant subspaces. Let us first consider the exponential type invariant subspaces with their corresponding non-linear time-fractional DCW equation \eqref{cdeqn}.
\subsection{Exact solutions of \eqref{cdeqn} using the exponential subspaces}
Now, let us first consider one-dimensional exponential linear space
\begin{equation}\label{l1}
{_3^1\mathcal{\hat{W}}_{1}}=\text{Span}\left\{e^{-(a_{10}x_1+a_{20}x_2+a_{30}x_3)}\right\},
\end{equation}
 which is invariant with respect to the given non-linear partial differential operator $\mathcal{\hat{H}}[u]$ if\\
 $P_1(u)=\left({-\sum\limits_{i=2}^{3}\frac{a_{i0}^2\tau_{i1}}{a_{10}^2}+\sum\limits_{i=1}^{3}\frac{a_{i0}\rho_{i1}}{2a_{10}^2}}\right)u+\tau_{10}$,
 $P_2(u)=\tau_{21}u+\tau_{20}$, $P_3(u)=\tau_{31}u+\tau_{30}$ and \\
 $Q_j(u)=\rho_{j1}u+\rho_{j0}$, $j=1,2,3$, because
\begin{equation*}\label{eg1invc}
\mathcal{\hat{H}}[\delta e^{-(a_{10}x_1+a_{20}x_2+a_{30}x_3)}] =\left( \sum_{i=1}^{3} a_{i0}^{2}\tau_{i0}-\sum_{i=1}^{3} a_{i0}\rho_{i0}\right)\delta e^{-(a_{10}x_1+a_{20}x_2+a_{30}x_3)}\in{{_3^1\mathcal{\hat{W}}_{1}}},\ \delta\in\mathbb{R}.
\end{equation*}
The above-mentioned case was listed in table \ref{q1} of case 1. For this case, the given equation \eqref{cdeqn} reduces into
\begin{eqnarray}\label{eg1}\begin{aligned}
\dfrac{\partial^{\alpha}u}{\partial t^{\alpha}}=
\mathcal{\hat{H}}[u]
\equiv &
 \dfrac{\partial}{\partial x_1}\left[
 \left(
 \left[{-\sum\limits_{i=2}^{3}\frac{a_{i0}^2\tau_{i1}}{a_{10}^2}+\sum\limits_{i=1}^{3}\frac{a_{i0}\rho_{i1}}{2a_{10}^2}}\right]u+\tau_{10}
 \right)\dfrac{\partial u}{\partial x_1}\right]
 +
\dfrac{\partial}{\partial x_2}\left[\left(\tau_{21}u+\tau_{20}\right)\dfrac{\partial u}{\partial x_2}\right]\\&
+\dfrac{\partial}{\partial x_3}\left[\left(\tau_{31}u+\tau_{30}\right)\dfrac{\partial u}{\partial x_3}\right]
+\sum\limits^{3}_{i=1}\left(\rho_{i1}u+\rho_{i0}\right)\left(\dfrac{\partial u}{\partial x_i}\right),\alpha\in(0,2]
\end{aligned}
\end{eqnarray}
with the initial conditions
\begin{eqnarray}
\begin{aligned}\label{in1}
&(i)\ u(x_1,x_2,x_3,0)= \varphi(x_1,x_2,x_3)\ \text{if}\ 0<\alpha\leq1,\\
&(ii)\ u(x_1,x_2,x_3,0)=\varphi(x_1,x_2,x_3)\ \&\ {\dfrac{\partial u}{\partial t}}\big{|}_{t=0}=\phi(x_1,x_2,x_3)\ \text{if}\ 1<\alpha\leq2.
\end{aligned}
\end{eqnarray}
Thus, we look for the exact solution of \eqref{eg1} as
\begin{equation}\label{eq1solnFDE}
u(x_1,x_2,x_3,t)= \Psi(t)e^{-(a_{10}x_1+a_{20}x_2+a_{30}x_3)}.
\end{equation}
Substitute \eqref{eq1solnFDE} in \eqref{eg1}, we obtain
\begin{equation}\label{re}
 \dfrac{d^\alpha \Psi(t)}{d t^\alpha} =\left( \sum\limits_{i=1}^{3} a_{i0}^{2}\tau_{i0}-\sum\limits_{i=1}^{3} a_{i0}\rho_{i0}\right)\Psi(t).
\end{equation}
We know that the Laplace transformation of the Caputo arbitrary-order derivative of $G(y)$ is obtained as \cite{kai,pi} 
 \[ L\left[\frac{d^\alpha G(y)}{dy^\alpha}\right]=s^\alpha L[G(y)]-\sum_{m=0}^{n-1}s^{\alpha-(m+1)}G^{(m)}(0),\ Re(s)>0,\ \alpha\in(n-1, n],\ n\in\mathbb{N}, \]
 where $G^{(m)}(0)= \dfrac{d^m G(y)}{dy^m}\mid_{y=0},\ y\in [0,\infty)$ and $L[G(y)]=\tilde{G}(s)=\int\limits_{0}^{\infty}e^{-st}G(r)dr.$\\
Let us first consider the integer-order cases $\alpha=1$ and $\alpha=2$. Hence the exact solutions of \eqref{eg1} associated with the given linear space \eqref{l1} are obtained as follows
 \begin{equation}\label{iofs}
u(x_1,x_2,x_3,t) = \left\{
\begin{array}{ll}
\gamma e^{A t-(a_{10}x_1+a_{20}x_2+a_{30}x_3)}, \, \text{if}\, \alpha =1,\\
\left(\gamma cosh(\sqrt{A}t)+ \dfrac{\beta}{\sqrt{A}} sinh(\sqrt{A}t)\right)e^{-(a_{10}x_1+a_{20}x_2+a_{30}x_3)}, \, \text{if} \,\alpha=2,
\end{array}\right.
\end{equation}
where $\gamma,\beta \in\mathbb{R} $ and  $A=\left( \sum\limits_{i=1}^{3} a_{i0}^{2}\tau_{i0}-\sum\limits_{i=1}^{3} a_{i0}\rho_{i0}\right).$
 Now, let $\alpha \in (0,1]$. Applying the Laplace transformation on both sides of the equation \eqref{re}, we get
 \begin{eqnarray}
 \begin{aligned}
& L\left[ \dfrac{d^\alpha \Psi(t)}{d t^\alpha} \right]=L\left[A \Psi(t)\right]& \text{ which gives}
\label{re1}\quad s^{\alpha}\tilde{\Psi}(s)-s^{\alpha-1}\Psi(0)=A \tilde{\Psi}(s),
  \end{aligned}
 \end{eqnarray}
 where $\Psi(0)=\gamma$ and $A=\left( \sum\limits_{i=1}^{3} a_{i0}^{2}\tau_{i0}-\sum\limits_{i=1}^{3} a_{i0}\rho_{i0}\right).$
Rearranging the terms and taking inverse Laplace transformation on both sides of the equation \eqref{re1}, we obtain
\begin{equation}
\Psi(t)=\gamma E_{\alpha,1}(A t^\alpha),\ \alpha\in(0,1],
\end{equation}
where $E_{\alpha,r}(t^\alpha)$ is the Mittag-Leffler function of two parameters which is defined as \cite{kai,mathai}
\[E_{\alpha,r}(t^\alpha)=\sum\limits_{m=0}^{\infty}\dfrac{(t^\alpha)^m}{\Gamma(m\alpha+r)}.\]
Next, we consider the case $\alpha \in (1,2]$. For this case, we again apply the Laplace transformation on both sides of the equation \eqref{re} which gives
\begin{eqnarray}
 \begin{aligned}
& L\left[ \dfrac{d^\alpha \Psi(t)}{d t^\alpha} \right]=L\left[A\Psi(t)\right],\,\,
\text{implying}\, \, s^{\alpha}\tilde{\Psi}(s)-s^{\alpha-1}\Psi(0)-s^{\alpha-2}\Psi'(0)=A\tilde{\Psi}(s).
  \end{aligned}
 \end{eqnarray}
Rearranging the terms and taking inverse Laplace transform on both sides of the last of the above equation, we obtain
\begin{equation}\label{psi}
\Psi(t)=\gamma E_{\alpha,1}(A t^\alpha)+ \beta t E_{\alpha,2}(A t^\alpha),\ \alpha\in(1,2],
\end{equation}
where $ \gamma=\Psi(0)$ and $\beta=\Psi'(0)$.
Hence the obtained exact solution of \eqref{eg1} is as follows,
\begin{equation}\label{fs}
u(x_1,x_2,x_3,t) = \left\{
\begin{array}{ll}
\gamma E_{\alpha,1}(A t^\alpha)e^{-(a_{10}x_1+a_{20}x_2+a_{30}x_3)}, \quad \text{if}\ \alpha \in (0,1],\\
\left(\gamma E_{\alpha,1}(A t^\alpha)+ \beta t E_{\alpha,2}(A t^\alpha)\right)e^{-(a_{10}x_1+a_{20}x_2+a_{30}x_3)}, \quad \text{if}\ \alpha \in (1,2],
\end{array}\right.
\end{equation}
where $\gamma=\Psi(0)$, $\beta=\Psi'(0)$ and $A=\left( \sum\limits_{i=1}^{3} a_{i0}^{2}\tau_{i0}-\sum\limits_{i=1}^{3} a_{i0}\rho_{i0}\right)
$.\\
Note that the solutions \eqref{fs} satisfy the given initial conditions \eqref{in1}. Further,
\begin{itemize}
\item[(a)] At $t=0$, $E_{\alpha,1}( 0)=1$. Thus, we obtain $u(x_1,x_2,x_3,0)= \gamma e^{-(a_{10}x_1+a_{20}x_2+a_{30}x_3)}$, if $\alpha\in(0,1]$.
\item[(b)] Now, let $\alpha\in(1,2]$. At $t=0$, we obtain
\begin{eqnarray*}
&u(x_1,x_2,x_3,0)= \gamma e^{-(a_{10}x_1+a_{20}x_2+a_{30}x_3)}\ \&\  \dfrac{\partial u}{\partial t}\big{|}_{t=0}= \beta e^{-(a_{10}x_1+a_{20}x_2+a_{30}x_3)}, \, \text{because}\\
 & E_{\alpha,1}( 0)=1, \dfrac{ d}{dt}\left(E_{\alpha,1}(A t^{\alpha})\right)=t^{\alpha-1}E_{\alpha,\alpha}(A t^\alpha) \,\, \text{and} \,\, \dfrac{ d}{dt}\left(tE_{\alpha,2}(A t^{\alpha})\right)=E_{\alpha,1}(A t^{\alpha}).
\end{eqnarray*}
\end{itemize}
From the above-obtained conditions, we get $ \varphi(x_1,x_2,x_3)=\gamma e^{-(a_{10}x_1+a_{20}x_2+a_{30}x_3)}$ and
$\phi(x_1,x_2,x_3)=\beta e^{-(a_{10}x_1+a_{20}x_2+a_{30}x_3)}.$
Additionally, it should be noted that when $\alpha=1$ and $\alpha=2$ and using the following properties of the Mittag-Leffler function \cite{mathai}
\begin{equation}\label{id}
E_{1,1}(y)=e^y ,E_{2,1}(y^2)=cosh(y)\,\text{and} \,E_{2,2}(y^2)=\dfrac{sinh(y)}{y} ,
\end{equation}
the obtained solutions \eqref{fs} are exactly same as the obtained integer-order solutions \eqref{iofs}.
\begin{nt}
We would like to point out when $\alpha\in(0,1]$, the given equation \eqref{eg1} is known as the time-fractional diffusion-convection equation while it is the diffusion-convection-wave equation if $\alpha\in(1,2]$. For these two cases, we derived the exact solutions \eqref{fs} of the given equation \eqref{eg1} along with appropriate initial conditions.
\end{nt}
Now, we consider the $(3+1)$-dimensional quadratic non-linear time-fractional diffusion-convection-wave equation as follows. For $\alpha\in(0,2],  $  we have
\begin{eqnarray}\label{eg2}\begin{aligned}
\dfrac{\partial^{\alpha}u}{\partial t^{\alpha}}= &
 \dfrac{\partial}{\partial x_1}\left[\left(\dfrac{\rho_{11}}{a_{10}}u+\tau_{i0}\right)\dfrac{\partial u}{\partial x_1}\right]+\sum\limits^{3}_{i=2}\tau_{i0}\dfrac{\partial^2 u}{\partial x_i^2}+
 \left(\rho_{11}u+\rho_{10}\right)\left(\dfrac{\partial u}{\partial x_1}\right)
+\sum\limits^{3}_{i=2}\rho_{i0}\left(\dfrac{\partial u}{\partial x_i}\right)
\end{aligned}
\end{eqnarray}
with the appropriate initial conditions
\begin{eqnarray}
\begin{aligned}
&(i)\ u(x_1,x_2,x_3,0)= \varphi(x_1,x_2,x_3)\  \text{if}\ 0<\alpha\leq1,\\
&(ii)\ u(x_1,x_2,x_3,0)=\varphi(x_1,x_2,x_3)\ \&\ {\dfrac{\partial u}{\partial t}}\big{|}_{t=0}=\phi(x_1,x_2,x_3)\ \text{if}\ 1<\alpha\leq2.
\end{aligned}
\end{eqnarray}
For the given equation \eqref{eg2}, we obtain the differential operator in the form
 $$\mathcal{\hat{H}}[u]
\equiv
 \dfrac{\partial}{\partial x_1}\left[\left(\dfrac{\rho_{11}}{a_{10}}u+\tau_{i0}\right)\dfrac{\partial u}{\partial x_1}\right]+\sum\limits^{3}_{i=2}\tau_{i0}\dfrac{\partial^2 u}{\partial x_i^2}
 +
 \left(\rho_{11}u+\rho_{10}\right)\left(\dfrac{\partial u}{\partial x_1}\right)
+\sum\limits^{3}_{i=2}\rho_{i0}\left(\dfrac{\partial u}{\partial x_i}\right) $$
which admits the linear space 
 \begin{eqnarray}\begin{aligned}\label{l2}
 ^1_3\mathcal{\hat{W}}_4=\text{Span}\{ e^{-a_{10}x_1},e^{-(a_{10}x_1+a_{21}x_2)},e^{-(a_{10}x_1+a_{31}x_3)},e^{-(a_{10}x_1+a_{21}x_2+a_{31}x_3)}\},\end{aligned}
\end{eqnarray}
because
\begin{eqnarray*}
\begin{aligned}
\mathcal{\hat{H}}[u]&=\mathcal{\hat{H}}[\delta_1e^{-a_{10}x_1}+\delta_2e^{-(a_{10}x_1+a_{21}x_2)}+\delta_3e^{-(a_{10}x_1+a_{31}x_3)}+\delta_4e^{-(a_{10}x_1+a_{21}x_2+a_{31}x_3)}]\\
&=\left(\tau_{10}a_{10}^2-\rho_{10}a_{10}\right)\delta_1e^{-a_{10}x_1}+\left(\tau_{10}a_{10}^2-\rho_{10}a_{10}+\tau_{20}a_{21}^2-\rho_{20}a_{21}\right)\delta_2e^{-(a_{10}x_1+a_{21}x_2)}\\
&+\left(\tau_{10}a_{10}^2-\rho_{10}a_{10}+\tau_{20}a_{21}^2-\rho_{20}a_{21}\tau_{30}a_{31}^2-\rho_{30}a_{31}\right)\delta_4e^{-(a_{10}x_1+a_{21}x_2+a_{31}x_3)}\\
&+\left(\tau_{10}a_{10}^2-\rho_{10}a_{10}+\tau_{30}a_{31}^2-\rho_{30}a_{31}\right)\delta_3e^{-(a_{10}x_1+a_{31}x_3)}\in{^1_3\mathcal{\hat{W}}_4}.
\end{aligned}
\end{eqnarray*}
Thus, we can write an exact solution of \eqref{eg2} as
\begin{equation}
u(x_1,x_2,x_3,t)=\sum_{q=1}^4\Psi_q(t)\xi_q(x_1,x_2,x_3),
\end{equation}
where the functions $\xi_1(x_1,x_2,x_3)=e^{-a_{10}x_1}$, $\xi_2(x_1,x_2,x_3)=e^{-(a_{10}x_1+a_{21}x_2)}$, $\xi_3(x_1,x_2,x_3)=e^{-(a_{10}x_1+a_{31}x_3)}$, $\xi_4(x_1,x_2,x_3)=e^{-(a_{10}x_1+a_{21}x_2+a_{31}x_3)}$ and the functions $\Psi_q(t)$ $(q=1,2,3,4)$ satisfy the following arbitrary-order ODEs,
\begin{equation}
\dfrac{d^{\alpha}\Psi_q(t)}{dt^\alpha}=A_q \Psi_q(t),\ q=1,2,3,4,
\end{equation}
where $ A_1=\tau_{10}a_{10}^2-\rho_{10}a_{10},$  $ A_2=\tau_{10}a_{10}^2-\rho_{10}a_{10}+\tau_{20}a_{21}^2-\rho_{20}a_{21},A_3=\tau_{10}a_{10}^2-\rho_{10}a_{10}+\tau_{30}a_{31}^2-\rho_{30}a_{31}$ and $A_4=\tau_{10}a_{10}^2-\rho_{10}a_{10}+\tau_{20}a_{21}^2-\rho_{20}a_{21}\tau_{30}a_{31}^2-\rho_{30}a_{31}$.\\
Note that for the integer-order cases $\alpha=1$ and $\alpha=2$, we obtain the solutions of \eqref{eg2} associated with the linear space \eqref{l2} as follows,
\begin{eqnarray}\label{iosol}
u(x_1,x_2,x_3,t) = \left\{
\begin{array}{ll} \sum\limits_{q=1}^4
\gamma_q e^{A_q t}\xi_q(x_1,x_2,x_3), \, \text{if} \,\alpha=1,\\
\sum\limits_{q=1}^4
\left[\gamma_q cosh(\sqrt{A_q}t)+ \dfrac{\beta_q}{\sqrt{A_q}}sinh(\sqrt{A_q}t)\right]\xi_q(x_1,x_2,x_3), \, \text{if} \, \alpha=2,
\end{array}\right.
\end{eqnarray}
where $\gamma_q,\beta_q\in \mathbb{R}.$ Proceeding as before, we obtain the exact solutions of \eqref{eg2} for $\alpha\in(0,2]$ in the form:
\begin{eqnarray}\label{sol}
u(x_1,x_2,x_3,t) = \left\{
\begin{array}{ll} \sum\limits_{q=1}^4
\gamma_q E_{\alpha,1}(A_q t^\alpha)\xi_q(x_1,x_2,x_3), \quad \alpha \in (0,1],\\
\sum\limits_{q=1}^4
\left[\gamma_q E_{\alpha,1}(A_q t^\alpha)+ \beta_q t E_{\alpha,2}(A_q t^\alpha)\right]\xi_q(x_1,x_2,x_3), \quad \alpha \in (1,2],
\end{array}\right.
\end{eqnarray}
where $\gamma_q=\Psi_q(0)$ and $\beta_q=\Psi_q'(0),q=1,2,3,4$.
Now, it is easy to check that the solutions \eqref{sol} of \eqref{eg2} satisfy the following initial conditions,
\begin{eqnarray}
\begin{aligned}
&(i)\ u(x_1,x_2,x_3,0)= \varphi(x_1,x_2,x_3) \ \text{if} \ 0<\alpha\leq1,\\
&(ii)\ u(x_1,x_2,x_3,0)=\varphi(x_1,x_2,x_3)\ \&\ {\dfrac{\partial u}{\partial t}}\big{|}_{t=0}=\phi(x_1,x_2,x_3)\ \text{if}\ 1<\alpha\leq2,
\end{aligned}
\end{eqnarray}
where $\varphi(x_1,x_2,x_3)=\sum\limits_{q=1}^4 \gamma_q\xi_q(x_1,x_2,x_3)$ and  $\phi(x_1,x_2,x_3)=\sum\limits_{q=1}^4 \beta_q\xi_q(x_1,x_2,x_3).$\\
 We also observe that when $\alpha=1$ and $\alpha=2$ and using the properties of the Mittag-Leffler function given in \eqref{id}, the solutions \eqref{sol} are exactly  the same as the integer-order solutions given in \eqref{iosol}.
\subsection{Exact solutions of \eqref{cdeqn} using the polynomial subspaces}
Here we consider the $(3+1)$-dimensional quadratic non-linear time-fractional DCW equation
\begin{eqnarray}\label{eg3}\begin{aligned}
\dfrac{\partial^{\alpha}u}{\partial t^{\alpha}}=
\mathcal{\hat{H}}[u]
\equiv &
\sum\limits^{3}_{r=1}
 \dfrac{\partial}{\partial x_r}\left[\left(\tau_{r1}u+\tau_{r0}\right)\dfrac{\partial u}{\partial x_r}\right]
+\sum\limits^{3}_{r=1}\rho_{r0}\left(\dfrac{\partial u}{\partial x_r}\right),\alpha\in(0,2]
\end{aligned}
\end{eqnarray}
along with the initial conditions
\begin{itemize}
\item[(i)] $u(x_1,x_2,x_3,0)= \varphi(x_1,x_2,x_3)$  if $\alpha\in(0,1]$,
\item[(ii)] $u(x_1,x_2,x_3,0)=\varphi(x_1,x_2,x_3)$ $\&$ ${\dfrac{\partial u}{\partial t}}\big{|}_{t=0}=\phi(x_1,x_2,x_3)$ if $\alpha \in(1,2]$.
\end{itemize}
The given non-linear partial differential operator $\mathcal{\hat{H}}[u]$ admits the 4-dimensional polynomial space
\begin{equation}
\label{l3}
^3_3\mathcal{\hat{W}}_4=\text{Span}\{1,x_1,x_2,x_3\},
\end{equation}
 which is listed in table \ref{q5} of case 20. When $\alpha=1$ and $\alpha=2$, the exact solutions of \eqref{eg3} associated with the linear space \eqref{l3} are as follows:
\begin{equation}\label{iofs1}
u(x_1,x_2,x_3,t) = \left\{
\begin{array}{ll}
\gamma_0 t +\gamma_1+\gamma_2x_1+\gamma_3x_2+\gamma_4 x_3
, \quad \text{if}\ \alpha =1,\\
\sum\limits_{i=1}^3 A_i\dfrac{t^{1+i}}{(1+i)!}+\gamma_1+ t\beta_1+\sum\limits_{i=1}^3\left(\gamma_{i+1}+t\beta_{i+1}\right)x_i, \quad \text{if}\ \alpha=2,
\end{array}\right.
\end{equation}
where $A_1=\sum\limits_{i=1}^3 \left(\tau_{i1}\gamma_{i+1}^2+\rho_{i0}\gamma_{i+1}\right)$, $A_2=\sum\limits_{i=1}^3 \left(2\tau_{i1}\gamma_{i+1}\beta_{i+1}+\rho_{i0}\beta_{i+1}\right)$ and $A_3=\sum\limits_{i=1}^3 \tau_{i1}\beta_{i+1}^2,\\\gamma_0,\gamma_i,\beta_i\in \mathbb{R},i=1,2,3,4.$
Proceeding as before, we obtain the exact solutions of the given equation \eqref{eg3} as follows:
\begin{equation}\label{fs1}
u(x_1,x_2,x_3,t) = \left\{
\begin{array}{ll}
\gamma_0 \dfrac{t^{\alpha}}{\Gamma{(\alpha+1)}} +\gamma_1+\gamma_2x_1+\gamma_3x_2+\gamma_4 x_3
, \quad \text{if}\ \alpha \in (0,1],\\
\sum\limits_{i=1}^3 A_i\dfrac{t^{\alpha+i-1}}{\Gamma(\alpha+i)}+\gamma_1+ t\beta_1+\sum\limits_{i=1}^3\left(\gamma_{i+1}+t\beta_{i+1}\right)x_i, \quad \text{if}\ \alpha \in (1,2].
\end{array}\right.
\end{equation}
 We wish to point out that the obtained exact solutions satisfy the given appropriate initial conditions. Additionally, we observe that $\varphi(x_1,x_2,x_3)=\gamma_1+\gamma_2x_1+\gamma_3x_2+\gamma_4 x_3  $ and $\phi(x_1,x_2,x_3)=\beta_1+\beta_2x_1+\beta_3x_2+\beta_4 x_3.$
It should be noted that when $\alpha=1$ and $\alpha=2$ and using the properties of gamma function, the obtained solutions \eqref{fs1} are exactly the same as the integer-order solutions \eqref{iofs1}.
\subsection{Exact solutions of \eqref{cdeqn} using the combinations of trigonometric and polynomial subspace}
Let us consider the 5-dimensional linear space
 \begin{equation}
 \label{l4}
 ^3_3\mathcal{\hat{W}}_5=\text{Span}\{\xi_{n} (x_1,x_2,x_3):n=1,2,3,4,5\}=\text{Span}\{1,{x_1},x_2,sin(\sqrt{a_{31}}x_3),cos(\sqrt{a_{31}}x_3)
\},
\end{equation}
which is invariant under $\mathcal{\hat{H}}[u]$ given in \eqref{cdop} along with $P_i(u)=\tau_{i1}u+\tau_{i0}$, $i=1,2$, $P_3(u)=\tau_{30}$, $Q_i(u)=\rho_{i0}$, $(i=1,2)$ and $Q_3(u)=0$. This case is listed in table \ref{q1} of case 3 (additionally, we choose $\rho_{11}=\rho_{21}=\rho_{30}=0$). Thus, the given equation \eqref{cdeqn} is reduced into
\begin{eqnarray}\label{eg4}\begin{aligned}
\dfrac{\partial^{\alpha}u}{\partial t^{\alpha}}=
\mathcal{\hat{H}}[u]
\equiv &
\sum\limits_{r=1}^{2}
 \dfrac{\partial}{\partial x_r}\left[\left(\tau_{r1}u+\tau_{r0}\right)\dfrac{\partial u}{\partial x_r}\right]+
 \tau_{30}\dfrac{\partial^2 u}{\partial x_3^2}
+\sum\limits^{2}_{r=1}\rho_{r0}\left(\dfrac{\partial u}{\partial x_r}\right),\alpha\in(0,2]
\end{aligned}
\end{eqnarray}
with the appropriate initial conditions
\begin{itemize}
\item[(i)] $u(x_1,x_2,x_3,0)= \varphi(x_1,x_2,x_3)$  if $0<\alpha\leq1$,
\item[(ii)] $u(x_1,x_2,x_3,0)=\varphi(x_1,x_2,x_3)$ $\&$ ${\dfrac{\partial u}{\partial t}}\big{|}_{t=0}=\phi(x_1,x_2,x_3)$ if $1<\alpha\leq2$.
\end{itemize}
For the integer-order cases $\alpha=1\ \text{and}\ \alpha=2$, we obtain the exact solutions of \eqref{eg4} associated with the linear space $^3_3\mathcal{\hat{W}}_5 $ as
\begin{eqnarray}
\begin{aligned}\label{iosol1}
u(x_1,x_2,x_3,t) =&
A_1t +\gamma_1+\gamma_2{x_1}+\gamma_3x_2\\&
+e^{-\tau_{30} t}\left[\gamma_4 sin(\sqrt{a_{31}}x_3)+\gamma_5cos(\sqrt{a_{31}}x_3)
\right], \, \text{if}\, \alpha =1,\\
u(x_1,x_2,x_3,t) =&
\sum\limits_{i=1}^3 A_i\dfrac{t^{i+1}}{(1+i)!}+\gamma_1+ t\beta_1+
\left(\gamma_{2}+t\beta_{2}\right)x_1
+\left(\gamma_{3}+t\beta_{3}\right)x_2
\\&+
cosh(\sqrt{-\tau_{30}} t)\left[\gamma_4 sin(\sqrt{a_{31}}x_3)+\gamma_5cos(\sqrt{a_{31}}x_3)
\right]
\\&+
\dfrac{sinh(\sqrt{-\tau_{30}} t)}{\sqrt{-\tau_{30}}}\left[\beta_4 sin(\sqrt{a_{31}}x_3)+\beta_5cos(\sqrt{a_{31}}x_3)
\right],
  \, \text{if}\, \alpha =2,
 \end{aligned}
\end{eqnarray}
where $A_1=\sum\limits_{i=1}^{2}\left(\tau_{i1}\gamma_{i+1}^2+\rho_{i0}\gamma_{i+1}\right)$, $A_2=\sum\limits_{i=1}^2\left(2\tau_{i1}\gamma_{i+1}\beta_{i+1}+\rho_{i0}\beta_{i+1}\right)$ and $A_3=\beta_2^2+\beta_3^2,\gamma_i,\beta_i\in \mathbb{R},i=1,2,3,4,5. $ \\
Proceeding as before, we obtain the exact solutions of \eqref{eg4} in the form
\begin{eqnarray}
\begin{aligned}\label{sol1}
u(x_1,x_2,x_3,t) =&
A_1\dfrac{t^{\alpha}}{\Gamma{(\alpha+1)}} +\gamma_1+\gamma_2{x_1}+\gamma_3x_2\\&
+E_{\alpha,1}(-\tau_{30} t^\alpha)\left[\gamma_4 sin(\sqrt{a_{31}}x_3)+\gamma_5cos(\sqrt{a_{31}}x_3)
\right], \quad \text{if}\ \alpha \in (0,1],\\
u(x_1,x_2,x_3,t) =&
\sum\limits_{i=1}^3 A_i\dfrac{t^{\alpha+i-1}}{\Gamma(\alpha+i)}+\gamma_1+ t\beta_1+
\left(\gamma_{2}+t\beta_{2}\right)x_1
+\left(\gamma_{3}+t\beta_{3}\right)x_2
\\&+
E_{\alpha,1}(-\tau_{30} t^\alpha)\left[\gamma_4 sin(\sqrt{a_{31}}x_3)+\gamma_5cos(\sqrt{a_{31}}x_3)
\right]
\\&+
t E_{\alpha,2}(-\tau_{30} t^\alpha)\left[\beta_4 sin(\sqrt{a_{31}}x_3)+\beta_5cos(\sqrt{a_{31}}x_3)
\right],
 \quad \text{if}\ \alpha \in (1,2].
 \end{aligned}
\end{eqnarray}
Also, it is easy to check that the solutions \eqref{sol1} satisfy the given appropriate initial conditions. In addition, we observe that $\varphi(x_1,x_2,x_3)=\sum\limits_{i=1}^5 \gamma_i\xi_i(x_1,x_2,x_3)$ and $\phi(x_1,x_2,x_3)=\sum\limits_{i=1}^5 \beta_i\xi_i(x_1,x_2,x_3).$
We also observe that when $\alpha=1$ and $\alpha=2$ and using the properties of gamma and the Mittag-Leffler functions, the above-obtained solutions \eqref{sol1} are exactly the same as the integer-order solutions \eqref{iosol1}.
\subsection{Exact solutions of \eqref{cdeqn} using the combinations of exponential, trigonometric and polynomial subspace}
Finally, we consider the $(3+1)$-dimensional quadratic non-linear time-fractional diffusion-convection-wave equation
\begin{eqnarray}\label{eg5}\begin{aligned}
\dfrac{\partial^{\alpha}u}{\partial t^{\alpha}}=
\mathcal{\hat{H}}[u]
\equiv &
\tau_{10}\dfrac{\partial^2 u}{\partial x_1^2}+
 \dfrac{\partial}{\partial x_2}\left(\left[\tau_{21}u+\tau_{20}\right]\dfrac{\partial u}{\partial x_2}\right)+
 \tau_{30}\dfrac{\partial^2 u}{\partial x_3^2}
+\sum\limits^{2}_{i=1}\rho_{i0}\left(\dfrac{\partial u}{\partial x_i}\right),\alpha\in(0,2]
\end{aligned}
\end{eqnarray}
with the initial conditions
\begin{itemize}
\item[(i)] $u(x_1,x_2,x_3,0)= \varphi(x_1,x_2,x_3)$  if $\alpha\in(0,1]$,
\item[(ii)] $u(x_1,x_2,x_3,0)=\varphi(x_1,x_2,x_3)$ $\&$ ${\dfrac{\partial u}{\partial t}}\big{|}_{t=0}=\phi(x_1,x_2,x_3)$ if $\alpha \in(1,2]$,
\end{itemize}
which admits the 5-dimensional linear space
\begin{equation}
\label{l5}
^3_3\mathcal{\hat{W}}_5=\text{Span}\{\xi_{n} (x_1,x_2,x_3):n=1,2,3,4,5\}=\text{Span}\{1,e^{-a_{11}x_1},x_2,sin(\sqrt{a_{31}}x_3),cos(\sqrt{a_{31}}x_3)
\}.
\end{equation}
 This case is listed in table \ref{q3} of case 15 (additionally, we can choose $\rho_{21}=\rho_{30}=0$)).
 For the integer-order cases $\alpha=1$ and $\alpha=2$, we obtain the exact solutions of \eqref{eg5} associated with the linear space $ ^3_3\mathcal{\hat{W}}_5$ as follows:
\begin{eqnarray}\begin{aligned}\label{iosol2}
u(x_1,x_2,x_3,t) =&
A_1 t +\gamma_1+\gamma_2e^{At}e^{-a_{11}x_1}+\gamma_3x_2\\&
+e^{-\tau_{30}t}\left[\gamma_4 sin(\sqrt{a_{31}}x_3)+\gamma_5cos(\sqrt{a_{31}}x_3)
\right], \, \text{if}\,\alpha =1,\\
u(x_1,x_2,x_3,t) =&
\sum\limits_{i=1}^3 A_i\dfrac{t^{i+1}}{(1+i)!}+\gamma_1+ t\beta_1+\left[\gamma_2 cosh(\sqrt{A}t)+ \dfrac{\beta_2}{\sqrt{A}} sinh(\sqrt{A}t)\right]
e^{-a_{11}x_1}\\&
+\left(\gamma_{3}+t\beta_{3}\right)x_2+
cosh(\sqrt{-\tau_{30}}t)\left[\gamma_4 sin(\sqrt{a_{31}}x_3)+\gamma_5cos(\sqrt{a_{31}}x_3)
\right]
\\&+
\dfrac{sinh(\sqrt{-\tau_{30}}t)}{\sqrt{-\tau_{30}}} \left[\beta_4 sin(\sqrt{a_{31}}x_3)+\beta_5cos(\sqrt{a_{31}}x_3)
\right],
 \, \text{if}\, \alpha =2.
 \end{aligned}
\end{eqnarray}
where $A=\left(\tau_{10}a_{11}^2-\rho_{10}a_{11}\right)$, $A_1=\left(\tau_{21}\gamma_3^2+\rho_{20}\gamma_3\right)$, $A_2=\left(2\tau_{21}\gamma_3\beta_3+\rho_{20}\beta_3\right)$ and $A_3=\beta_3^2,\gamma_i,\beta_i\in \mathbb{R},i=1,2,3,4,5.$
For the arbitrary-order case, we obtain the exact solutions of \eqref{eg5} as follows,
\begin{eqnarray}\begin{aligned}\label{sol2}
u(x_1,x_2,x_3,t) =&
A_1\dfrac{t^{\alpha}}{\Gamma{(\alpha+1)}} +\gamma_1+\gamma_2E_{\alpha,1}(A t^\alpha)e^{-a_{11}x_1}+\gamma_3x_2\\&
+E_{\alpha,1}(-\tau_{30} t^\alpha)\left[\gamma_4 sin(\sqrt{a_{31}}x_3)+\gamma_5cos(\sqrt{a_{31}}x_3)
\right], \quad \text{if}\ \alpha \in (0,1],\\
u(x_1,x_2,x_3,t) =&
\sum\limits_{i=1}^3 A_i\dfrac{t^{\alpha+i-1}}{\Gamma(\alpha+i)}+\gamma_1+ t\beta_1+
\left[\gamma_2E_{\alpha,1}(A t^\alpha)+t\beta_2E_{\alpha,2}(A t^\alpha) \right]e^{-a_{11}x_1}\\&
+\left(\gamma_{3}+t\beta_{3}\right)x_2+
E_{\alpha,1}(-\tau_{30} t^\alpha)\left[\gamma_4 sin(\sqrt{a_{31}}x_3)+\gamma_5cos(\sqrt{a_{31}}x_3)
\right]
\\&+
t E_{\alpha,2}(-\tau_{30} t^\alpha)\left[\beta_4 sin(\sqrt{a_{31}}x_3)+\beta_5cos(\sqrt{a_{31}}x_3)
\right],
 \quad \text{if}\ \alpha \in (1,2].
 \end{aligned}
\end{eqnarray}
 Also, it can be checked that the solutions \eqref{sol2} satisfy the given appropriate initial conditions. Additionally, we find that $ \varphi(x_1,x_2,x_3)=\sum\limits_{i=1}^5 \gamma_i\xi_i(x_1,x_2,x_3)$ and $\phi(x_1,x_2,x_3)=\sum\limits_{i=1}^5 \beta_i\xi_i(x_1,x_2,x_3).$
We also note that when $\alpha=1$ and $\alpha=2$ and using the properties of the Mittag-Leffler and gamma functions, the solutions \eqref{sol2} are exactly the same as the integer-order solutions \eqref{iosol2}.

\begin{figure}[h!]
\begin{subfigure}{0.5\textwidth}
 \includegraphics[width=8cm, height=6cm]{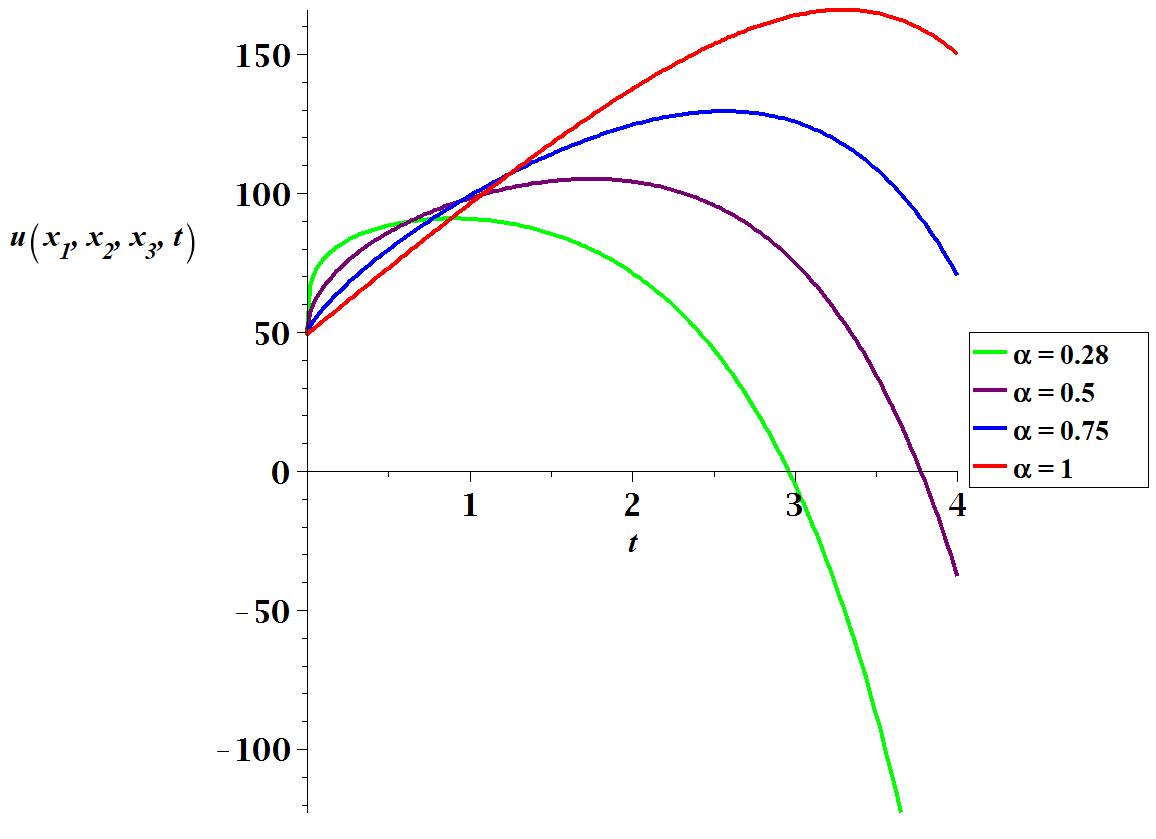}
    \caption{{ (t,u)-plane ($x_1=4,x_2=5,x_3=8$)}}
    \label{fig:2d1}\end{subfigure}
    \begin{subfigure}{0.5\textwidth}
    \includegraphics[width=8cm, height=6.5cm]{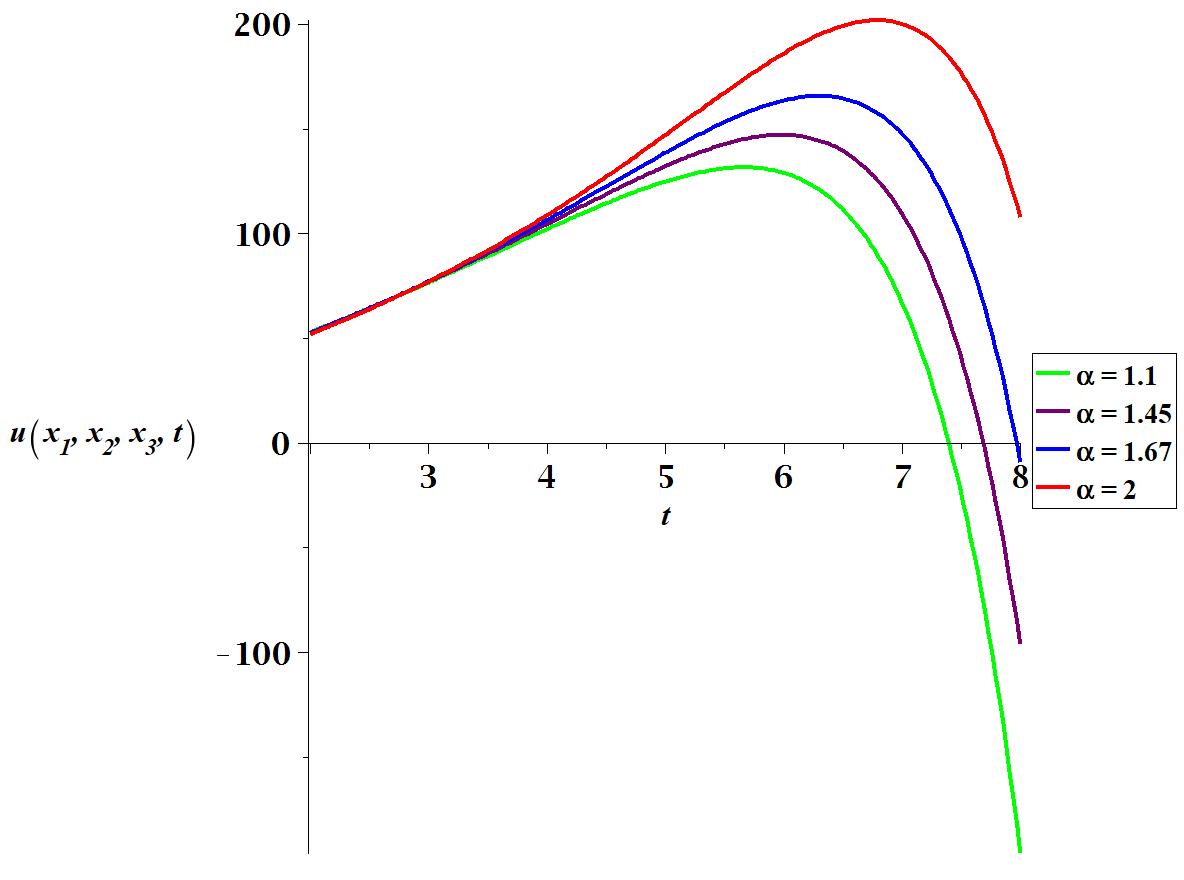}
    \caption{{ (t,u)-plane ($ x_1=4,x_2=5,x_3=8$) }}
    \label{fig:2d2}
    \end{subfigure}
    \caption{{2D graphical representations of solutions \eqref{sol2} for different values of $\alpha$.}}
        \label{fig1}
\end{figure}

\begin{figure}[h!]
\begin{subfigure}{0.5\textwidth}
 \includegraphics[width=9cm, height=6.5cm]{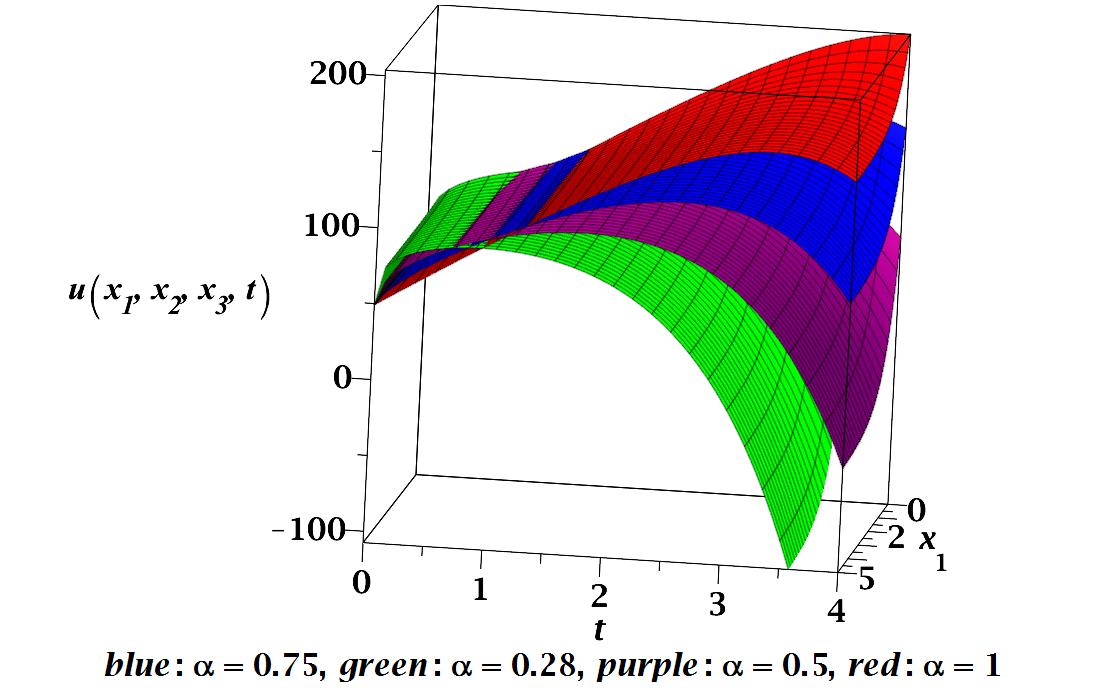}
    \caption{{$x_2=5,x_3=8$}}
    \label{fig:3d1}
    \end{subfigure}
    \begin{subfigure}{0.5\textwidth}
    \includegraphics[width=8cm, height=6.5cm]{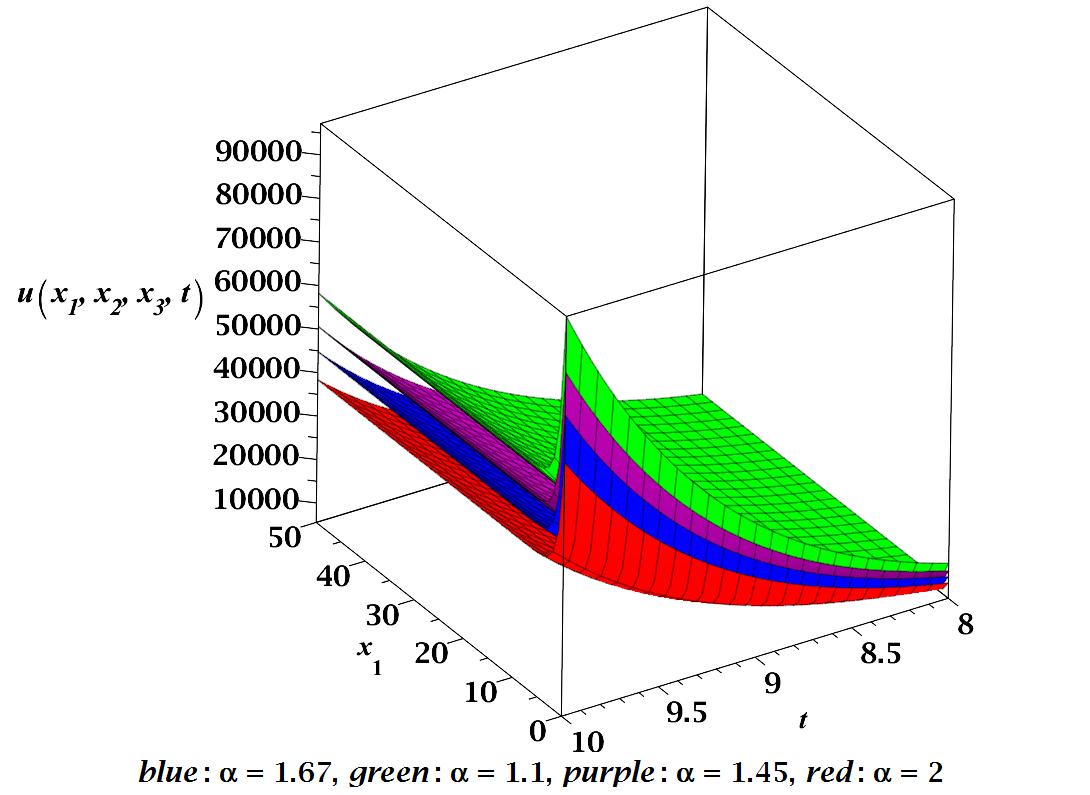}
    \caption{{ $ x_2=5,x_3=8$}}
    \label{fig:3d4}
    \end{subfigure}
    \caption{{3D graphical representations of solutions \eqref{sol2} for different values of $\alpha$ in $ (x_1,t,u)$-space.}}
        \label{fig2}
\end{figure}

\begin{figure}[h!]
    \begin{subfigure}{0.5\textwidth}
    \includegraphics[width=8cm, height=6.5cm]{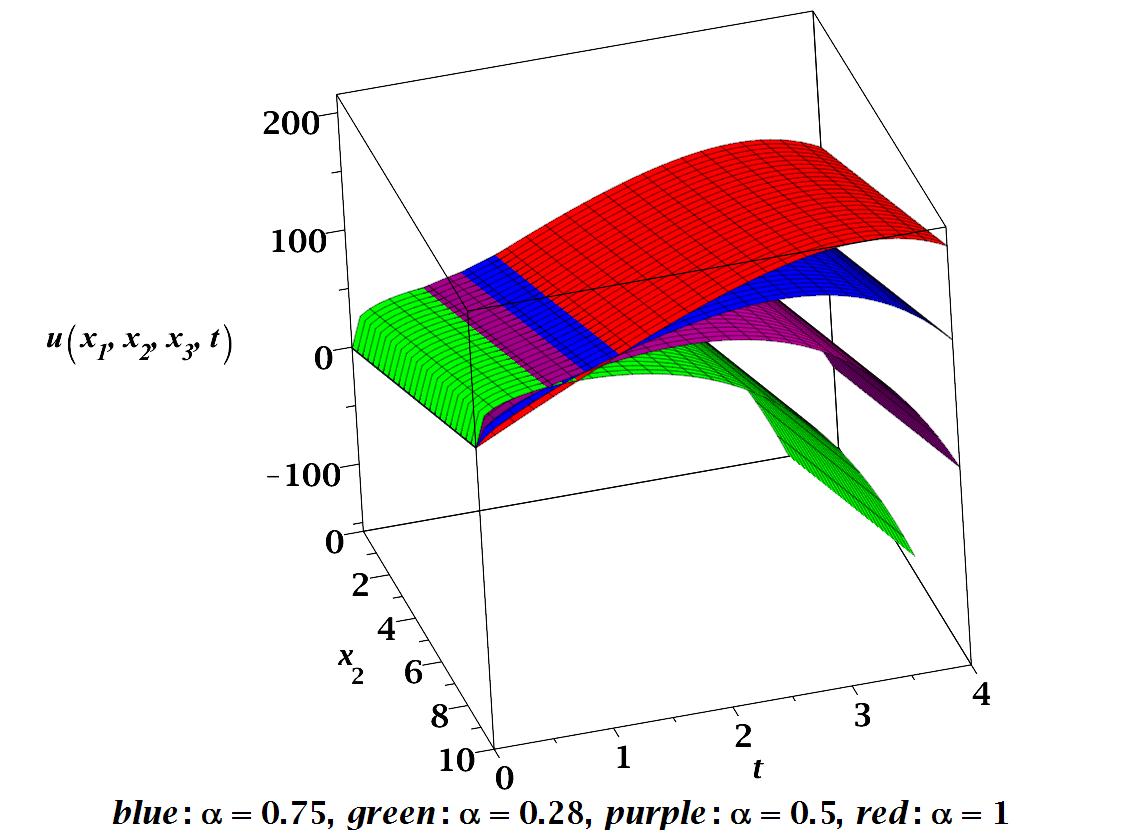}
    \caption{{  $ x_1=4,x_3=8$}}
    \label{fig:3d2}
    \end{subfigure}
    \begin{subfigure}{0.5\textwidth}
    \includegraphics[width=8cm, height=6.5cm]{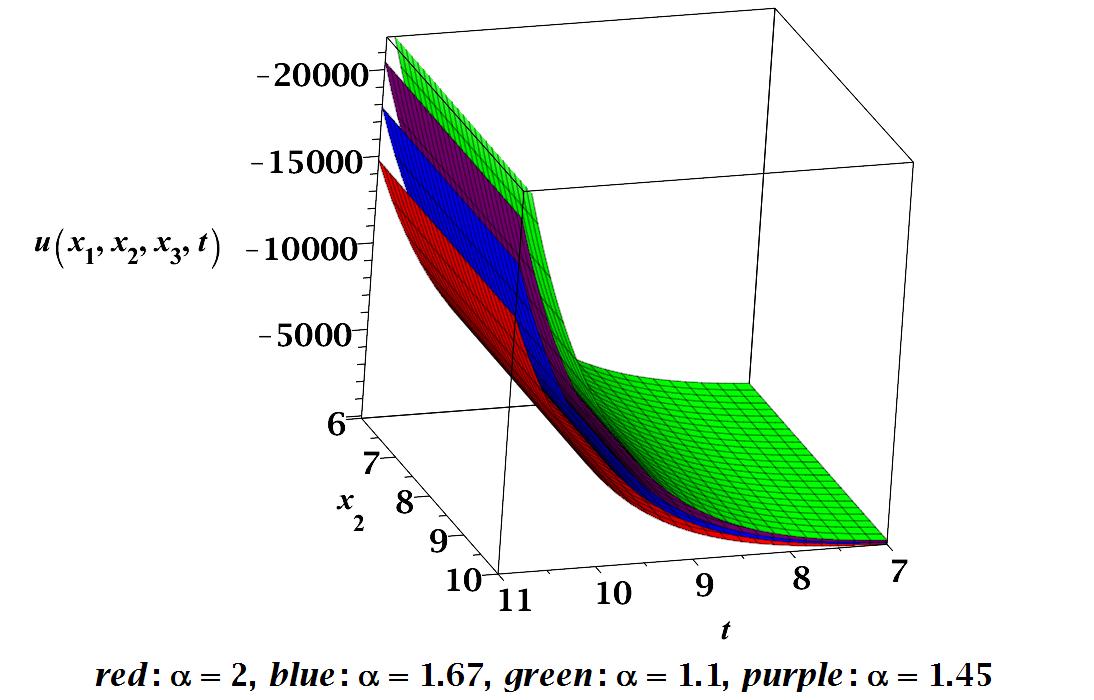}
    \caption{{   $ x_1=4,x_3=8$}}
    \label{fig:3d5}
    \end{subfigure}
    \caption{{3D graphical representations of solutions \eqref{sol2} for different values of $\alpha$ in $ (x_2,t,u)$-space.}}
        \label{fig3}
\end{figure}

\begin{figure}[h!]
    \begin{subfigure}{0.5\textwidth}
    \includegraphics[width=8cm, height=6.5cm]{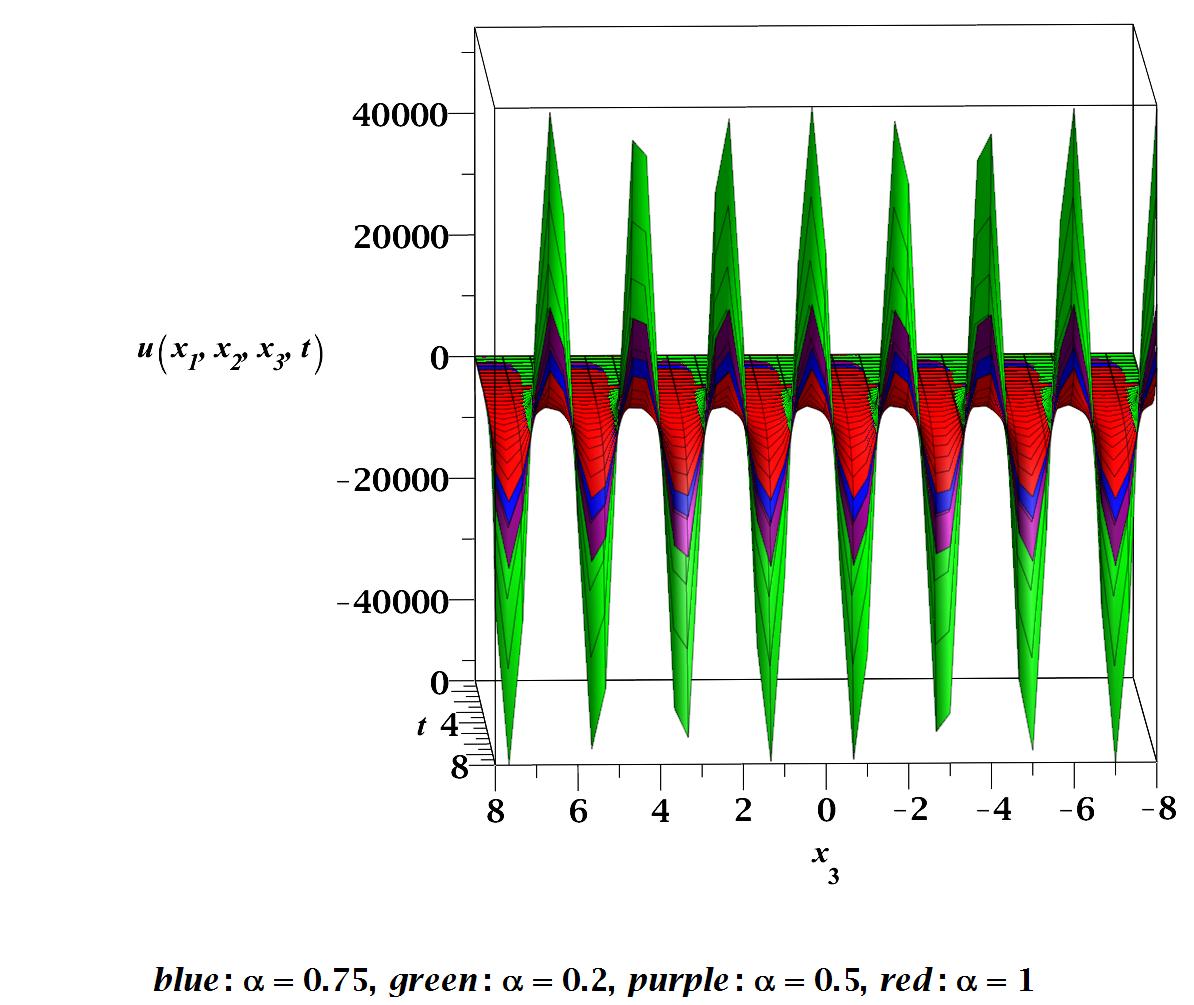}
    \caption{{  $ x_1=4,x_2=5$}}
        \label{fig:3d3}
\end{subfigure}
    \begin{subfigure}{0.5\textwidth}
        \includegraphics[width=9cm, height=6.5cm]{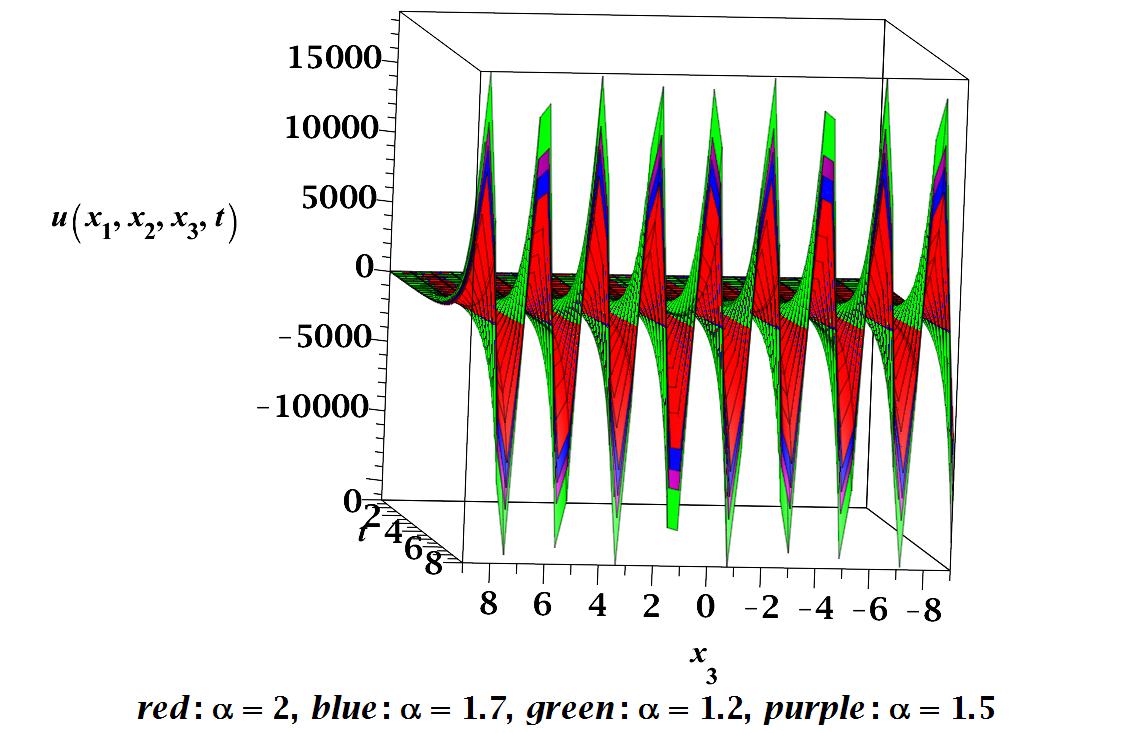}
    \caption{{  $ x_1=4,x_2=5$}}
    \label{fig:3d6}
    \end{subfigure}
    \caption{{3D graphical representations of solutions \eqref{sol2} for different values of $\alpha$ in $(x_3,t,u)$-space.}}
        \label{fig4}
\end{figure}

 In order to elucidate the dynamic behavior of the obtained solutions \eqref{sol2}, we give the 2-dimensional (2D) and 3-dimensional (3D) graphical representations of the solutions for various values of $\alpha \in(0,2]$ under certain parametric values, $A_1=50,\gamma_1=\gamma_2=a_{11}=1,\gamma_3=10,\gamma_4=3,\gamma_5=2 $ which are shown from Figure \ref{fig1} to Figure \ref{fig4}.
\section{Discussions and concluding remarks}{
Finally, a critical discussion is provided on the implications of the invariant subspace method developed above to various types of integer and fractional order diffusion equations and their modifications in different spatial dimensions. Also, brief concluding remarks are presented.
\subsection{Discussions}
In this subsection, we briefly summarize the applicability of the invariant subspace method to several well known classical diffusion type equations and their modifications in order to emphasize the relevance of the general method derived in the main paper.
We know that the well-known classical diffusion equation has the form
\begin{equation}\label{111}
\dfrac{\partial u}{\partial t}=k\dfrac{\partial^2u}{\partial x^2},\ k\in\mathbb{R}.
\end{equation}
The water movement in soil has been a topic of discussion from time immemorial and its practical significance is unquestionable. Even after many years after its introduction, soil water modelling as an extension of Darcy's law is invoking great curiosity among many scientists. The $(1+1)$-dimensional vertical flow leads to the formulation of the following integer-order $(1+1)$-dimensional convection-diffusion equation, which can be viewed as a combination of the equation of continuity for conservation of water mass and the Buckingham-Darcy's law for unsaturated flow \cite{thesis},
\begin{eqnarray}\label{11}
\begin{aligned}
&\dfrac{\partial u}{\partial t} = \dfrac{\partial }{\partial x}& \left(A(u)\dfrac{\partial u}{\partial x} \right)-&\dfrac{d B(u)}{d u}\dfrac{\partial u}{\partial x},\\
& & \uparrow_\text{capillarity}  & \uparrow_\text{gravity}
\end{aligned}
\end{eqnarray}
where $u=u(x,t)$ gives the volumetric water content for a given time $t$ at the depth of $x$ from the surface of the soil and the functions $A(u) \, \&\, B(u) $ provide the  concentration-dependent soil-water diffusivity and conductivity, respectively. Moreover, $\dfrac{\partial }{\partial x} \left(A(u)\dfrac{\partial u}{\partial x} \right)$ represents the effect of capillarity  and $ \dfrac{d B(u)}{d u}\dfrac{\partial u}{\partial x} $ represents the effect of gravity in the movement of water in the soil, because
\begin{itemize}
\item[(i)] Buckingham-Darcy's law for unsaturated flow: $F =-A(u)\dfrac{\partial u}{\partial x} + B(u)$.
\item[(ii)] Continuity for conservation of water mass equation: $\dfrac{\partial u}{\partial t}=-\dfrac{\partial F}{\partial x}$.
\end{itemize}
Equation \eqref{11} also describes many non-linear physical situations that occur in different situations like semiconductor diffusivity, porous medium flow and so on.
Exact solutions of \eqref{11} have been investigated in \cite{thesis,li}. In the literature, the $(1+1)$-dimensional generalized non-linear convection-reaction-diffusion equation \cite{ps1,gs} reads as follows:
\begin{equation}\label{13}
\dfrac{\partial u}{\partial t}=\dfrac{\partial }{\partial x} \left(A(u)\dfrac{\partial u}{\partial x} \right)+C(u)\dfrac{\partial u}{\partial x}+R(u),
\end{equation}
where the functions $A(u)$, $C(u)$ and $R(u)$ denote the phenomena of the diffusion, convection and reaction, respectively.  Exact solutions of \eqref{13} have been derived through the invariant subspace method in \cite{ps1}.  Galaktionov and Svirshchevskii \cite{gs} have investigated the exact solutions of specific types of \eqref{13} using the invariant subspaces that are given below.
\begin{itemize}
\item[(i)] When $A(u)=u^\sigma$, $C(u)=0$ and $R(u)=-u^{1-\sigma}$, then the above equation \eqref{13} is reduced into
\begin{equation}\label{14}
\dfrac{\partial u}{\partial t}=\dfrac{\partial }{\partial x} \left(u^\sigma\dfrac{\partial u}{\partial x} \right)-u^{1-\sigma},\ \sigma>0,
\end{equation}
which is known as famous diffusion-absorbtion equation. The above equation \eqref{14} admits exact solution based on the two-dimensional linear space $\mathcal{\hat{W}}_2=\text{Span}\left\{1,x\right\}$ along with the pressure transformation $v=u^\sigma$.
\item[(ii)] If $A(u)=u^{-\frac{4}{3}}$, $C(u)=0$ and $R(u)=u^{\frac{7}{3}}$, then the equation \eqref{13} becomes a reaction-diffusion equation
\begin{equation}
 \dfrac{\partial u}{\partial t}=\dfrac{\partial }{\partial x} \left(u^{-\frac{4}{3}}\dfrac{\partial u}{\partial x} \right)+u^{\frac{7}{3}},
\end{equation}
which admits the exact solution associated with the five-dimensional linear space $\mathcal{\hat{W}}_5=\text{Span}\left\{1,x,x^2,x^3,x^4\right\}$ along with the pressure transformation $v=u^{-\frac{4}{3}}$.
\item[(iii)] If $A(u)=u^{-\frac{4}{3}}$, $C(u)=0$ and $R(u)=-u^{-\frac{1}{3}}$, then the equation \eqref{13} is reduced into
\begin{equation}\label{15}
 \dfrac{\partial u}{\partial t}=\dfrac{\partial }{\partial x} \left(u^{-\frac{4}{3}}\dfrac{\partial u}{\partial x} \right)-u^{-\frac{1}{3}},
\end{equation}
which is called a reaction-absorbtion equation. The above equation \eqref{15} admits an exact solution based on the five-dimensional linear space
$$ \mathcal{\hat{W}}_5=\text{Span}\{1,sin(\dfrac{2}{\sqrt{3}}x),sin(\dfrac{4}{\sqrt{3}}x),cos(\dfrac{2}{\sqrt{3}}x),cos(\dfrac{4}{\sqrt{3}}x)\},$$
along with the pressure transformation $v=u^{-\frac{4}{3}}$.
\item[(iv)] When $A(u)=u^{-\frac{3}{2}}$, $C(u)=0$ and $R(u)=0$, then the equation \eqref{13} becomes a fast diffusion equation
 \begin{equation}\label{o15}
 \dfrac{\partial u}{\partial t}=\dfrac{\partial }{\partial x} \left(u^{-\frac{3}{2}}\dfrac{\partial u}{\partial x} \right),
\end{equation}
which admits an exact solution based on the five-dimensional linear space $\mathcal{\hat{W}}_5=\text{Span}\left\{1,x,x^2,x^3,x^4\right\}$ along with the pressure transformation $v=u^{-\frac{3}{2}}$.
\end{itemize}
The most general class of $(n+1)$-dimensional non-linear convection-reaction-diffusion equation \cite{CRD2} reads in the form
\begin{eqnarray}
\begin{aligned}\label{gcrd}
\dfrac{\partial u}{\partial t} = \bigtriangledown \cdot\left(A(u)\bigtriangledown u\right)+B(u)\cdot\bigtriangledown u+ C(u),
\end{aligned}
\end{eqnarray}
where $ u=u(x_1,x_2,\dots,x_n,t)$,
$ \bigtriangledown=\left(\dfrac{\partial}{\partial x_1},\dfrac{\partial}{\partial x_2},\dots, \dfrac{\partial}{\partial x_n}\right),$ and
 `$\cdot$' means scalar product. Note that Lie symmetries of $(2+1)-$dimensional convection-reaction-diffusion equation have been discussed in \cite{CRD2}.
Also, we wish to point out that exact solutions of $(2+1)$-dimensional convection-diffusion equation have been investigated through the Lie symmetry method in \cite{CRD2,thesis,cdeq}.
\\
It should be noted that the formatting of the classical diffusion equation \eqref{111} is based on the consideration of mean square displacement which is linear and time-dependent (ie., $<x^2(t)>\sim k t$).
Here we wish to point out that non-integer order derivatives are used to model  anomalous diffusion, particularly kinetic equations of fractional order have been investigated in \cite{ahan1,ahan3,book1,lenzi,vitali,ralf,ginoa} which is useful in the context of anomalous diffusion. Anomalous diffusion can be characterized by an asymptotic long-time non-linear behaviour of the mean square displacement which reads as
\begin{equation}
<x^2(t)>\sim k t^\alpha,\ \alpha\in(0,2],\ t\rightarrow\infty ,
\end{equation}
and which can be classified into two different classes that are (i) sub-diffusion if $0<\alpha\leq1$ and (ii) super-diffusion if $1<\alpha\leq2$. Thus, the governing equation of anomalous diffusion equation reads as follows
\begin{equation}\label{112}
\dfrac{\partial^\alpha u}{\partial t^\alpha}=k\dfrac{\partial^2u}{\partial x^2},\ 0<\alpha\leq2, k\in\mathbb{R}.
\end{equation}
Metzler and Klafter \cite{ralf} have investigated that diffusion equation of fractional order describes a non-Markovian diffusion process with a memory. Ginoa et al \cite{ginoa} have discussed that diffusion equation of fractional-order describing relaxation phenomena in complex viscoelastic materials.}\\
{
In the literature, the invariant subspace method for the $(1+1)$-dimensional non-linear TFPDEs has been discussed in
\cite{ru2,ra,pra4,pa1,s1,pp1,pp2,ru3,praf,gara1,gara2}.
Let us consider the $(1+1)$-dimensional non-linear TFPDE
 \begin{equation}\label{d11}
\dfrac{\partial^\alpha u}{\partial t^\alpha}=\mathbf{H}[u],\alpha>0,
\end{equation}
where $u=u(x,t),x\in \mathbb{R},t\geq0$ and $\mathbf{H}[u]$ is the sufficiently smooth $k$-th order non-linear differential operator, that is $\mathbf{H}[u]=\mathbf{H}\left(x,u,\frac{\partial u}{\partial x},\dots,\frac{\partial^k u}{\partial x^k}\right)$, $k\in\mathbb{N}.$
Now, we define the $n$-dimensional linear space $ \mathcal{W}_n=\text{Span}\{\xi_q(x) \big{|} q=1,2,\dots,n\},$ where $\{\xi_q(x)\}_{q=1}^n$ is a linearly independent set. 
Suppose that $ \mathcal{W}_n(n<\infty)  $ is invariant under the given differential operator $\mathbf{H}[u].$
Thus, we have
 \begin{equation}\label{11invc}
 \mathbf{H}\left[\sum\limits_{q=1}^n \kappa_q\xi_q(x)\right]=\sum\limits_{q=1}^n\mathcal{K}_q\left(\kappa_1,\dots,\kappa_n\right)\xi_q(x),\kappa_q\in \mathbb{R},
 \end{equation}
  where the functions $\mathcal{K}_q\left(\kappa_1,\dots,\kappa_n\right) (q=1,2,\dots,n) $ are the coefficients of expansion with respect to the basis set $\{\xi_q(x)\}_{q=1}^n.$
The above equation \eqref{11invc} can be written as $\mathbf{H[u]}\in \mathcal{W}_n$ whenever $u\in\mathcal{W}_n$.
 Then there exists an exact solution for the given equation \eqref{d11} in the following finite separable form,
\[u(x,t)=\sum_{q=1}^n\Psi_q(t)\xi_q(x)\]
if the functions $ \Psi_q(t)$ satisfy the system of fractional-order ODEs
\[ \dfrac{d^\alpha \Psi_q(t)}{d t^\alpha}= \Lambda_q(\Psi_1(t),\dots,\Psi_n(t)),q=1,2,\dots,n.
\]
We know that the $(1+1)$-dimensional generalized time-fractional reaction-convection-diffusion-wave equation reads as follows,
\begin{equation}
\dfrac{\partial^\alpha u}{\partial t^\alpha}=\dfrac{\partial }{\partial x} \left(A(u)\dfrac{\partial u}{\partial x} \right)+C(u)\dfrac{\partial u}{\partial x}+R(u),\alpha\in(0,2],
\label{crd11}
\end{equation}
where $u=u(x,t),x\in\mathbb{R},t>0,$ the functions $A(u),C(u) $ and $R(u)$ as mentioned above in equation \eqref{gcrd}. The exact solutions of \eqref{crd11} with various choices of functions  $A(u),C(u) $ and $R(u)$ have been studied by using the invariant subspace method \cite{ra,pra4,gara1,pa1}.
Additionally, the discussed method was developed for $(2+1)$-dimensional  non-linear TFPDEs in \cite{ppa}. Also, we would like to point out that the exact solutions of $(2+1)$-dimensional time-fractional reaction-convection-diffusion-wave equation
\begin{equation}\label{kk}
	\frac{\partial^{\alpha}u}{\partial t^{\alpha}}=
 \sum\limits^{2}_{r=1} \frac{\partial }{\partial x_r}\left(A_r(u)\frac{\partial u}{\partial x_r}\right)+ \sum\limits^{2}_{r=1}B_r(u)\frac{\partial u}{\partial x_r}+C(u),\ 0<\alpha\leq2,
 \end{equation}
 along with appropriate initial conditions
 \begin{itemize}
\item[(i)] $u(x_1,x_2,0)= \varphi(x_1,x_2)$  if $0<\alpha\leq1$,
\item[(ii)] $u(x_1,x_2,0)=\varphi(x_1,x_2)$ \& ${\dfrac{\partial u}{\partial t}}\big{|}_{t=0}=\phi(x_1,x_2)$ if $1<\alpha\leq2$,
\end{itemize}
  have been derived through the invariant subspace method in \cite{ppa}. }
\subsection{Concluding remarks}
In this work, we have developed a generalization of the invariant subspace method for $(m+1)$-dimensional non-linear TFPDEs for the first time. More specifically, the efficacy and applicability of the method are systematically investigated through the $(3+1)$-dimensional non-linear time-fractional diffusion-convection-wave equation along with appropriate initial conditions.  This systematic investigation has provided a method as to how to find a large class of various types of linear subspaces with different dimensions for the underlying non-linear TFPDEs. Additionally, we have shown that the obtained invariant subspaces help one to derive a variety of exact solutions that can be expressed as the combinations of exponential, trigonometric, polynomial and well-known Mittag-Leffler functions.
 Also, we feel that the obtained results and concepts of the method for the $(m+1)$-dimensional non-linear TFPDEs have not been discussed in the existing literature. We know that due to the violation of the standard properties by fractional-order derivatives, no well-defined analytical methods exist to analyse them systematically. So the above discussed systematic procedure will help one to find exact solutions of various types of scalar and coupled non-linear fractional-order PDEs in future. Hence these investigations show that the invariant subspace method is very effective and powerful mathematical tool to derive exact solutions of higher dimensional scalar and coupled non-linear fractional PDEs along with appropriate initial conditions.
 \section*{Acknowledgements}
 The work of M. L. is supported by a DST-SERB National Science Chair.


\begin{thebibliography}{aaaa}
\bibitem{pi} I. Podlubny, Fractional Differential Equations, Academic Press, New York, 1999.
\bibitem{kai} K. Diethelm, The Analysis of Fractional Differential Equations, Springer, Berlin, 2010.
\bibitem{kts} A.A. Kilbas, J.J. Trujillo, H.M. Srivastava, Theory and Applications of Fractional Differential Equations, Elseiver, Amsterdam, 2006.
\bibitem{vas5}  V.E. Tarasov, Fractional Dynamics: Applications of Fractional Calculus to Dynamics of Particles, Fields and Media, Nonlinear Physical Science, Springer, Heidelberg, Germany, 2011.
 \bibitem{mo11} H.G. Sun, Y. Zhang, D. Baleanu, W. Chen, Y.Q. Chen, A new collection of real world applications of fractional
calculus in science and engineering, Commun. Nonlinear Sci. Numer. Simulat. 64(2018) 213-231.
\bibitem {mn}{ F. Mainardi, Fractional Calculus: Some Basic Problems in Continuum and Statistical Mechanics. In: A. Carpinteri, F. Mainardi, (eds.) Fractals and Fractional Calculus in Continuum Mechanics, Springer, Vienna, 1997,  pp. 291-348 doi: 10.1007/978-3-7091-2664-6\_7.}
%
\bibitem{vas7} V.E. Tarasov, J.J. Trujillo, Fractional power-law spatial dispersion in electrodynamics, Annals of Physics 334(2013) 1-23.
\bibitem{vas14} C. Ionescu, A. Lopes, D. Copot, J.A.T. Machado, J.H.T. Bates, The role of fractional calculus in modeling biological phenomena: A review, Commun. Nonlinear Sci. Numer. Simul.  51(2017) 141-159.
\bibitem {btr} R.L. Bagley, P.J. Torvik, On the appearance of the fractional derivative in the behavior of real materials, ASME J. Appl. Mech. 51(1984) 294-298.
\bibitem{vas2} V.E. Tarasov, Review of some promising fractional physical models, Internat. J. Modern Phys. B 27(2013) 1330005.
\bibitem{bb1} T. Bakkyaraj, R. Sahadevan, Group formalism of Lie transformations to time-fractional partial differential equations, Pramana-J. Phys. 85(2015) 849-860.
\bibitem{lk} S. Yu. Lukashchuk, Conservation laws for time-fractional subdiffusion and diffusion-wave equations, Nonlinear Dyn. 80(2015) 791-802.
\bibitem{CRD2} R. Cherniha, M. Serov, Y. Prystavka, A complete Lie symmetry classification of a class of
(1+2)-dimensional reaction-diffusion-convection equations, Commun. Nonlinear Sci. Numer. Simulat. 92(2021) 105466.
\bibitem{ps1} P. Prakash, New exact solutions of generalized convection-reaction-diffusion equation, Eur. Phys. J. Plus 134(2019) 261.
\bibitem{CRD4} T. Harko, M.K. Mak, Exact travelling wave solutions of non-linear reaction-convection-diffusion equations-An Abel equation based approach, J. Math. Phys. 56(2015) 111501.
\bibitem{ra} R. Sahadevan, T. Bakkyaraj, Invariant subspace method and exact solutions of certain nonlinear time fractional partial differential equations, Fract. Calc. Appl. Anal. 18(2015) 146-162.
\bibitem{pra4} R. Sahadevan, P. Prakash, Exact solution of certain time fractional nonlinear partial differential equations, Nonlinear Dyn. 85(2016) 659-673.
\bibitem{pra3} R. Sahadevan, P. Prakash, Exact solutions and maximal dimension of invariant subspaces of time fractional coupled nonlinear partial differential equations, Commun. Nonlinear Sci. Numer. Simulat. 42(2017) 158-177.
\bibitem{ps2} S. Choudhary, P. Prakash, V. Daftardar-Gejji, Invariant subspaces and exact solutions for a system of fractional PDEs in higher dimensions, Comp. Appl. Math. 38(2019) 126.
\bibitem{ru2} R.K. Gazizov, A.A. Kasatkin, Construction of exact solutions for fractional order differential equations by invariant subspace method. Comput. Math. Appl. 66(2013) 576-584.
\bibitem{pa1} P. Artale Harris, R. Garra, Analytic solution of nonlinear fractional Burgers-type equation by invariant subspace method, Nonlinear Stud. 20(2013) 471-481.
\bibitem{s1} S. Choudhary, V. Daftardar-Gejji, Invariant subspace method: a tool for solving fractional partial differential equations, Fract. Calc. Appl. Anal. 20(2017) 477-493.
\bibitem{pp1} P. Prakash, Invariant subspaces and exact solutions for some types of scalar and coupled time-space fractional diffusion equations, Pramana-J. Phys. 94 (2020) 103(18p).
\bibitem{pp2} P. Prakash, S. Choudhary,  V. Daftardar-Gejji, Exact solutions of generalized nonlinear time-fractional reaction-diffusion equations with time delay, Eur. Phys. J. Plus 135(2020) 490(24p).
\bibitem{ru3} W. Rui, Idea of invariant subspace combined with elementary integral method for investigating exact solutions of time-fractional NPDEs, Appl. Math. Comput. 339(2018) 158-171.
\bibitem{pra2} R. Sahadevan, P. Prakash, On Lie symmetry analysis and invariant subspace methods of coupled time fractional partial differential equations, Chaos, Solitons and Fractals 104(2017) 107-120.
\bibitem{praf} P. Prakash, On group analysis, conservation laws and exact solutions of time-fractional Kudryashov-Sinelshchikov equation, Comput. Appl. Math 40(2021) 162.
\bibitem{kdv} R. Sahadevan, T. Bakkyaraj, Invariant analysis of time fractional generalized Burgers and Korteweg-de Vries equations, J. Math. Anal. Appl. 393(2012) 341-347.
\bibitem{ru4} R.K. Gazizov, A.A. Kasatkin, S.Yu. Lukashchuk, Symmetry properties of fractional diffusion equations, Phys. Scr. T136(2009) 014016 (5p). (doi:10.1088/0031-8949/
2009/T136/014016)
\bibitem{pra1} P. Prakash, R. Sahadevan, Lie symmetry analysis and exact solution of certain fractional ordinary differential equations, Nonlinear Dyn. 89(2017) 305-319.
\bibitem{pra5} R. Sahadevan, P. Prakash, Lie symmetry analysis and conservation laws of certain time fractional partial differential equations, Int. J. of Dynamical Systems and Differential Equations, 9(2019) 44-64.
\bibitem{nas} A.M. Nass, Lie symmetry analysis and exact solutions of fractional ordinary differential equations with neutral delay, Appl. Math. Comput. 347(2019) 370-380.
\bibitem{ppp} K. Sethukumarasamy, P. Vijayaraju, P. Prakash, On Lie symmetry analysis of certain coupled fractional ordinary differential equations, J. Nonlinear Math. Phys. 28(2021) 219-241.
\bibitem {Ge} V. Daftardar-Gejji, H. Jafari, Adomian decomposition: A tool for solving a system of fractional differential equations, J. Math. Anal. Appl. 301(2005) 508-518.
\bibitem{mo} S. Momani, Z. Odibat, Analytical solution of a time-fractional Navier-Stokes equation by Adomian decomposition method, Appl. Math. Comput. 177(2006) 488-494.
\bibitem{mo2} Z. Odibat, S. Momani, A generalized differential transform method for linear partial differential equations of fractional order, Appl. Math. Lett. 21(2008) 194-199.
\bibitem{mma} W.X. Ma, M.M. Mousa, M.R. Ali, Application of a new hybrid method for solving singular fractional Lane-Emden-type equations in astrophysics, Mod. Phys. Lett.B 34(2020) 2050049(10p).
\bibitem{gs} V.A. Galaktionov,  S.R. Svirshchevskii, Exact Solutions and Invariant Subspaces of Nonlinear Partial Differential Equations in Mechanics and Physics, Chapman and Hall/CRC, London, 2007.

\bibitem{gara1}  {P. Artale Harris}, R. Garra, Nonlinear heat conduction equations with memory: physical meaning and analytical results, J. Math. Phys. 58(2017) 063501.
\bibitem{gara2} R. Garra, Z. Tomovski, Exact results on some nonlinear Laguree-type diffusion equations, Math. Model. Anal. 26(2021) 72-81.
 {\bibitem {IVSM} A. H. A. Kader, M. S. A. Latif, D. Baleanu, Some exact solutions of a variable coefficients fractional biological population model, Math. Meth. Appl. Sci. 44(2021) 4701-4714.}
\bibitem{wx}  W.X. Ma, A refined invariant subspace method and applications to evolution equations, Sci. China Math. 55(2012) 1769-1778.
\bibitem{wy} W.X. Ma, Y. Liu, Invariant subspaces and exact solutions of a class of dispersive evolution equations, Commun. Nonlinear Sci. Numer. Simulat. 17(2012) 3795-3801.
\bibitem{wx1} W.X. Ma, Y. Zhang, Y. Tang, J. Tu, Hirota bilinear equations with linear subspaces of solutions, Appl. Math. Comput. 218(2012) 7174-7183.
\bibitem{yem} Y. Ye, W.X. Ma, S. Shen, D. Zhang, A class of third-order nonlinear evolution equations admitting invariant subspaces and associated reductions, J. Nonlinear Math. Phys. 21(2014) 132-148.
\bibitem{li} H. Liu, Invariant subspace classification and exact solutions to the generalized nonlinear D-C equation, Appl. Math. Lett. 83(2018) 164-168.
\bibitem{zq} C. Zhu, C. Qu, Invariant subspaces of the two-dimensional nonlinear evolution equations, symmetry 8(2016) 128.
\bibitem{ppa} P. Prakash, K.S. Priyendhu, K.M. Anjitha, { Initial value problem for the (2 + 1)-dimensional
time-fractional generalized convection–reaction–diffusion
wave equation: invariant subspaces and exact solutions, Comp. Appl. Math.
41(2022) 30.}
\bibitem{mathai} A.M. Mathai,  H.J. Haubold, Special Functions for Applied Scientists. Springer, New York, 2008.
%
%







\bibitem{thesis} M.P. Edwards, Exact solutions of nonlinear diffusion-convection equations, PhD thesis, University of Wollongong, 1997.

\bibitem{ahan1} A. Hanyga, Multidimensional solutions of space-fractional diffusion equations, Proc. R. Soc. Lond. A 457(2001) 2993-3005.
 \bibitem{ahan3} A. Hanyga, Multidimensional solutions of space-time-fractional diffusion equations, Proc. R. Soc. Lond. A 458(2002) 429-450.
 \bibitem{book1}L.R. Evangelista, E.K. Lenzi, Fractional Diffusion Equations and Anomalous Diffusion, Cambridge University Press, 2018.
 \bibitem{lenzi} E.K. Lenzi, H.V. Ribeiro, A.A. Tateishi, R.S. Zola, L.R. Evangelista, Anomalous diffusion and transport in heterogeneous systems separated by a membrane, Proc. R. Soc. A 472(2016) 20160502.
 \bibitem{vitali} S. Vitali, V. Sposini, O. Sliusarenko, P. Paradisi, G. Castellani, G. Pagnini, Langevin equation in complex media and anomalous diffusion, J. R. Soc. Interface 15(2018) 20180282.

\bibitem{cdeq} E. Demetriou, N.M. Ivanova, C. Sophocleous, Group analysis of (2+1)- and (3+1)-dimensional diffusion-convection equations, J. Math. Anal. Appl. 348(2008) 55-65.


\bibitem{ralf} R. Metzler, J. Klafter, The random walk's guide to anomalous diffusion: a fractional dynamics approach, Physics Reports 339(2000) 1-77.



\bibitem{ginoa} M. Ginoa, S. Cerbelli, H. E. Roman, Fractional diffusion equation and relaxation in complex viscoelastic materials, Physica A 191(1992) 449-453.




\bibitem{rui2022}{W. Rui, X. Yang, F. Chen,
Method of variable separation for investigating exact solutions and dynamical properties of the time-fractional Fokker-Planck equation,
Physica A 595(2022) 127068.}
\end{thebibliography}
\end{document}